\documentclass[aps,twocolumn,reprint,superscriptaddress]{revtex4}
\usepackage{amsmath}
\usepackage{mathtools}
\usepackage{graphicx}
\usepackage{amssymb}
\usepackage{amsfonts}
\usepackage{color}
\usepackage{mathrsfs}
\usepackage{gensymb}
\usepackage{textcomp}
\usepackage[capitalize]{cleveref}

\DeclareRobustCommand*{\rectangle}{{\setlength{\fboxsep}{-.5pt}\fbox{\phantom{l}}}}

\newcommand{\numberset}{\mathbb}

\newcommand{\Z}{\numberset{Z}}

\newcommand{\bx}{\mathbf{x}}
\newcommand{\bC}{\mathbf{C}}

\newcommand{\be}{\mathbf{e}}
\newcommand{\by}{\mathbf{y}}

\newcommand{\bF}{\mathbf{F}}

\newcommand{\bna}{\boldsymbol{\nabla}}

\usepackage{siunitx}

\newcommand{\di}{{i\mkern1mu}}

\begin{document}

\title{Inelastic  rotations and pseudo-turbulent plastic avalanches in crystals}%

\author{R. Baggio}
\affiliation{LSPM, CNRS UPR3407, Paris Nord Sorbonne Universit\'e, 93400, Villateneuse,  France}
\affiliation{PMMH, CNRS UMR 7636 ESPCI ParisTech, 10 Rue Vauquelin,75005, Paris, France}
\affiliation{UMR SPE 6134, Universit\'e de Corse, CNRS, Campus Grimaldi, 20250, Corte, France}
\author{O. U. Salman}
\affiliation{LSPM, CNRS UPR3407, Paris Nord Sorbonne Universit\'e, 93400, Villateneuse,  France}
\author{L. Truskinovsky}
\affiliation{PMMH, CNRS UMR 7636 ESPCI ParisTech, 10 Rue Vauquelin,75005, Paris, France}

\begin{abstract}
Plastic deformations in crystals   produce microstructures  with  randomly  oriented patches of   unstressed lattice forming  complex textures. We use a  novel mesoscopic Landau-type  tensorial model of crystal plasticity   to show that in such textures   rotations can originate from    crystallographically exact micro-slips which  self organize in the form of laminates of a pseudo-twin type. The formation of such laminates can be viewed as an effective internal 'wrinkling'  of the crystal lattice. While such 'wrinkling'  disguises  itself  as  an elastically neutral  rotation,  behind it  is inherently dissipative, dislocation-mediated process.   Our    numerical experiments  reveal   pseudo-turbulent effective rotations with power-law distributed spatial correlations which    suggests that 
  the process of dislocational self-organization  is inherently unstable and 
points towards the necessity of  a probabilistic description of crystal plasticity.

\end{abstract}
\maketitle

\section{Introduction}

The emerging experimental evidence of intermittent avalanches and scale-free dislocation patterns triggered a shift from macro- to micro-scale modeling efforts in crystal plasticity \cite{Zaiser2019-ti,Ovaska2017-cz,Weiss2015-eh}. The recorded temporal and spatial correlations were interpreted as evidence of complex cooperative dynamics at sub-continuum scales~\cite{Groma2019-gq,Papanikolaou2017-ld,Zhang2020-ax}, moreover, the observed hierarchically organized deformation fields were likened to scale-free turbulent   flows \cite{Cottrell2002-dh,Choi2012-tz,
 Bouchaud1995-pb,Radjai2002-cj,Odunuga2005-ap,Berdichevsky2019-rm}. 
 Since fluid turbulence largely relies on vortices, the question arises whether large rotations play a similarly crucial role in crystal plasticity \cite{Marcinkowski1989-ku,Kunin1990-fi,Shilko2016-cw,Andrade2005-dl,Kulagin2017-dy}.

\emph{Inelastic rotations.} When crystalline specimens are deformed plastically,  both elastic and inelastic rotations are  revealed through the emergence of deformation-induced crystallographic textures. They have been understood as a strain-accommodation strategy, allowing the crystal  to ensure lattice compatibility  without accumulation of  considerable elastic stresses.   \cite{Oddershede2015-lq,Hemery2019-he}. Energetically  neutral     rotations   are  also behind the localization of dislocations in wall structures  which separate misorientated lattice patches  \cite{Sethna2003-qz,Winther2008-hl,Dunne2012-cp,Castelluccio2017-xp,Vinogradov2018-gt,Arora2020-fi}. While the formulation of an adequate description of texture development is  commonly recognized as one of the most challenging tasks for theories of crystal plasticity,  the   microscopic mechanisms of particularly large  rotations involved in  textures formation remain obscure \cite{Asaro1983-du,Asaro1985-nm,Besseling1994-rr,Florando2006-ui,Chen2013-fh,Oddershede2015-lq,Hemery2019-he}
 
Plastic rotations in crystals  mainly  occur  by slip, however  
 other dislocation mediated modes of inelastic deformation may also contribute to changes in crystallographic texture  \cite{Van_der_Giessen1991-sz,Dafalias1990-gv,Dunne2012-cp,Wierzbanowski2011-kv,Dafalias1990-gv}. 
Thus, recent   molecular dynamic simulations of   single-crystal plasticity 
   suggested that, at least in the case of the discontinuous yield of defect-free crystals,  
 the patchy local reorientation of the crystal lattice is  due to deformation-induced micro- or nano-twinning \cite{ Zepeda-Ruiz2017-nd, Korchuganov2019-bw, Ko2020-bw,Bertin2020-hv,Wang2022-om}; 
 in physical experiments similar effects were also observed ~\cite{Barba2017-ad,Zhai2022-ar}.  
In this paper, we corroborate the idea of large inelastic rotations by micro-twinning theoretically while emphasizing the idea that such twinning ultimately relies  on dislocation glide. More specifically, we conduct numerical experiments  showing explicitly how crystallographically specific  inelastic deformations of twinning type can disguise themselves as large crystal rotations. 
 

\emph{Previous work.} Considerable efforts have been recently devoted to the task of small-scale modeling of crystal plasticity at a reasonable computational cost. A fully detailed description of plastic flows in crystals is possible only by molecular dynamics (MD) or molecular statics (MS) approaches \cite{Arsenlis2007-ke,Baruffi2019-qw,Zepeda-Ruiz2021-dw}, which accurately represent micro-mechanisms of plastic response while relying minimally on phenomenology. However, in most applications, such an approach is prohibitively computationally expensive, even if one deals with ultra-short time scales and ultra-small samples. 
 Partial  averaging of atomistic MD has emerged in the form of a continuum phase-field crystal method (PFC) \cite{Salvalaglio2020-eb,Chan2010-qt},  however  the resulting microscopically-detailed description of lattices still remains two demanding in terms of computer time, at least in the case of developed plastic flows with a realistic number of interacting dislocations.  
 
 The discrete dislocation dynamics (DDD)  approach was created to overcome the short-scale focus of atomistic methods and inform various classical continuum models. The DDD modeling is very powerful, allowing one, for instance, to model the organization of dislocations into cell structures \cite{Devincre2008-to,Ispanovity2014-lt,Gomez-Garcia2006-zn}. However, the DDD models contain many parameters since the processes of dislocation nucleation, interaction with defects, self-locking, climbing, etc., have to be prescribed phenomenologically through specific local rules coming from independent phenomenological constructs. Other major challenges in the DDD framework include   accounting for large plastic distortions   and  incorporating the effects of anisotropic elasticity
  \cite{Cai2006-fe}. Coarse graining of DDD has been attempted in the mesoscopic stochastic continuum dislocation dynamics (CDD)   where dislocation microstructures are modeled by various continuum dislocation density fields \cite{Starkey2020-qb}.  While various phenomenological closure relations have been proposed to model the evolution of such fields, the systematic development of this approach is hindered by the fact that rigorous statistical averaging in the ensemble of strongly interacting dynamic defects is still a tough challenge \cite{Monavari2016-fv,Lin2021-pq,Groma2021-eh}. 
  
  In search of a micro-macro compromise, a  multi-scale quasi-continuum   (QC) approach was developed based on the observation that a fully atomistic resolution is necessary only in small spatial regions, while in the rest of the computational domain, where the deformation fields change slowly, the classical continuum theory can be still used \cite{Shenoy1999-gh}.  The necessity of patching the continuum and discrete subdomains poses, however, a complex problem. Also, the necessity to resolve all the scales fully adequately makes this method hardly applicable to the description of collective dislocation processes.

A powerful mesoscale approach to crystal plasticity, which involves averaging over small scales and therefore allows one to treat dislocations in a continuum framework, is the phase field dislocation dynamics  (PFDD)  \cite{Javanbakht2016-dr,Bruzy2022-bg}.  In this approach, lattice slips are described by scalar order parameters, while the energy wells correspond to  lattice invariant shears. 
 The main problem of this approach is that lattice invariant shears are resolved by scalar order parameters only approximately even if   
 PFDD is extended to finite strains  \cite{Biscari2016-ie}. 
  
\emph{Mesoscopic tensorial model.}    All these computational platforms are successful in addressing particular time and length scales, with higher-scale models (more coarse-grained) relying on input from lower-scale models (less coarse-grained). However, as we have seen above,  each one of them has its limitations. 

To achieve a  compromise between more and less coarse-grained models we adopt  in our numerical experiments  a novel approach \cite{Salman2011-ij,Baggio2019-rs}, which is both versatile and synthetic in the above sense. Known as the mesoscopic tensorial model (MTM), it   represents  a crystal as a collection of homogeneously deforming elastic elements whose nonlinear elastic response is governed by globally periodic  potential defined in the space of metric tensors. The potential   is  designed to respect the geometrically nonlinear kinematics of the lattice,  as originally envisioned in  \cite{Ericksen1980-km,Boyer1989-hn,Folkins1991-em,Wang1993-nf,Waal1990-fd,Kaxiras1994-uf,Parry1998-sv, Conti2004-sv}. From the perspective of the ensuing   Landau-type  model, the elastic potential has   an infinite number of  equivalent energy wells, and therefore plastically deformed crystals can be viewed as coherent mixtures of equivalent  'phases' \cite{Baggio2019-rs,Salman2021-sn}.

To follow the response of the system one can use the real physical space or the configurational space of metric tensors where individual mesoscopic elements are represented by individual configurational points. As long as the  affine configuration remains  stable, the  points, representing different elastic elements   in the configurational space are all superimposed  and follow together the prescribed loading path.  After the instability, the configurational points spread around the configuration space with different  equivalent energy  wells getting  unequally populated. During such configurational spreading   the crystal is deforming plastically as  the elastic  energy is   lost  irreversibly. In particular, plastic yield of a pristine crystal can be  interpreted as a massive escape of the configurational points from the reference energy well while  plastic 'mechanisms' can be linked to the configuirational 'flows' along low-barrier valleys in the energy landscape. Friction-type dissipation, controlling dynamics in continuum crystal plasticity, emerges in such a theory as a result of a homogenized description of overdamped athermal dynamics in a rugged energy landscape which takes the form of a succession of  the  discontinuous branch switching events  \cite{Puglisi2005-lg}.  
   
The MTM based computational approach to crystal plasticity bears some resemblance to the local version of the quasi-continuum model \cite{Tadmor1996-qi,Li2004-vb,Hayes2005-ql} and also  has features in common with tensorial generalizations of   phase-field approaches to dislocational plasticity  \cite{Rodney2003-wy,Levitas2012-zx,Biscari2016-ie,Denoual2016-qm,Hu2021-mq}. Its main advantage is  the geometrically adequate account for both large strains and large rotations.  Another important benefit  is the possibility to model both short and long-range interactions of dislocations without any macroscopic phenomenological  assumptions. For instance, dislocation nucleation, dislocation annihilation, and dislocation locking emerge naturally from this modeling framework, moreover no ad-hoc rules are needed for specifying the activated slip planes. 

We reiterate that the most important assumption behind  MTM is that the crystal admits a pseudo-continuum strain energy density whose invariance is dictated  by the global symmetry group of simple lattices,  known as $GL(3,\mathbb{Z})$  \cite{Pitteri2002-rm}. From this point of view, the MTM can be considered as a multidimensional, finite-strain generalization of the   Peierls-Nabarro \cite{Peierls1940-ix,Nabarro2002-js} and Frenkel-Kontorova \cite{Frenkel1939-tk} one-dimensional models which also operate with globally periodic energies. However, in contrast to these purely prototypical models, the MTM potentials, constructed using ab initio methods, allow  one to make fully quantitative predictions about complex material response associated with plastic flows in crystals with particular crystallographic symmetries.

\emph{Benchmark test.} As a proof of principle,  we develop  in this paper a MTM description of a single plastic avalanche resulting from a brittle-like yield of a  homogeneously deformed pristine crystal. Our study was inspired by the recent fully atomistic simulations of bulk single-crystal plasticity in the uniaxially compressed  body-centered cubic  metal tantalum \cite{Zepeda-Ruiz2017-nd}. In the reported numerical experiments a defect-free, perfect crystal yielded discontinuously after reaching critical stress. Analysis of crystal configurations attained under strain revealed that the perfect metal yielded largely by deformation twinning — that is, by sudden strain-induced reorientation of the crystal lattice within bounded volumes of the material. To rationalize these observations we conducted our own numerical experiments using the novel MTM approach.  

In our numerical experiments we used   quasistatic  hard loading device (strain-controlled) and the response was obtained by  incremental   minimization of  the total  energy of the collection of coupled elastic elements.  In such apparently pseudo-elastic approach the energy  dissipation is taking place   during discontinuous   branch switching events  \cite{Puglisi2005-lg}. In the language of crystal plasticity  the branch switching  can be associated with   dislocation avalanches which involve collective nucleation, annihilation and reorganization of dislocations.

If a pristine crystal is loaded quasistatically, the very first avalanche of this type, signaling the catastrophic, brittle like  transition from affine to non-affine configuration,  takes the form of a massive system-size   dislocation  nucleation.  As we show, during such an event, multiple dislocations appear  collectively while  the lattice   rearranges itself into  almost fully relaxed domains (or patches)  with different orientations.  Our numerical experiments  provided a compelling evidence that in most of such domains, the apparent rigid rotations are achieved through coordinated, spatially distributed lattice invariant shears,    resulting in  the formation of highly specific micro-twinned  laminates. 

The importance  of energy-minimizing macroscale laminates  has been already realized  in continuum   crystal plasticity \cite{Ortiz1999-pp,Conti2015-ly}.  In this macroscopic framework  lamination was shown to result from the nonconvexity of the effective energy functional induced by geometric softening and/or latent hardening. Here we focus on the consequences of the microscopic  energy nonconvexity  associated with the presence of lattice invariant shears. We show that it can indeed lead to the formation of crystallographically specific laminates    at the microscale and that  the corresponding oscillatory band-like microstructures (representing alternating crystal orientations), can be interpreted as micro-twinning.

\emph{Main results.} In contrast to physical experiments,  the MTM modeling allows one to track the deformation history of individual elastic elements.  Using this capability of MTM,  we were able  to trace how exactly  the  micro-twin laminate  form and how they self-organize to present themselves  macroscopically as  pseudo-rigid rotations. The dissipative, dislocation-mediated nature of such self-organization suggests that at least some of the macroscale textures,   sometimes naively associated with purely elastic or even rigid rotations,      still emerge from the collective motion of dislocations.   In particular, using MTM we could study the transient process of avalanche unfolding in full detail and show that the modulated structure of a micro-twin results from the  propagation of an interface between the unstable homogeneous configuration and the stable oscillatory configuration. 

Furthermore, our numerical experiments focused on transient dynamics revealed that the formation of disoriented patches of the unstressed lattice during discontinuous avalanches is inherently unstable even though the sizes and the misorientations of the individual patches end up being highly correlated. The quantitative statistical analysis of the emerging spatial correlations suggests that the underlying process of dislocational self-organization indeed resembles  fluid turbulence.  All this  points towards the necessity of a probabilistic description of crystal plasticity, at least if the study of plastic fluctuations is at stake.
 
\emph{Organization of the paper. } The rest of the paper is structured as follows. In Section \ref{sec:themodel} we formulate the MTM approach, and  introduce the atomistically informed  energy density accounting for the global symmetry of a square lattice. The numerical  set up and the outcomes  of our numerical experiments are presented in Section \ref{sec:num}. In Section \ref{sec:ie} we reveal the mechanism  of rigid rotations recorded in our numerical experiments and link them to micro-twinning. In Section \ref{sec:pm} we present the analysis of the two  simplified models whose aim is to elucidate in the most transparent form the kinetic mechanism of dislocation driven  micro-twinning  disguised as rotation. The pseudo turbulent analogy is discussed in Section \ref{turb}. Finally, in the last Section \ref{sec:conc} we summarize  the  results and present our conclusions.

\section{The model}
\label{sec:themodel}
 
 In this paper  we limit our analysis to the simplest nontrivial 2D problem while assuming that   a model  crystal can be represented by  a collection of $N \times N$ mesoscopic triangular discrete  elements   organized in a  square lattice filling the macroscopic domain $\Omega_0$.  The internal scale of order $N^{-1}$ is then viewed as a physical parameter defining the  (Kolmogorov-type) cutoff beyond which the deformation is considered homogeneous.  To describe the deformation of  discrete elements we introduce  the piecewise smooth mapping $\by = \by(\bx)$, where $\by $ are their actual and  $\bx$ their  reference coordinates. We   then associate with each element an elastic energy density  $\phi$ which depends on the metric tensor $\bC=\bF^T\bF$,  where $ \bF=\bna\by $ is the deformation gradient.   
 
 \emph{Configurational space.} All deformations that map a  Bravais lattice into itself  will be accounted of if we   require that $\phi(\bC)=\phi(\bf m^T\bC \bf m)$ for  any  $\bf m$   from   an infinite  discrete group  $GL(2,\mathbb{Z}) = \left\{{\bf m},\, m_{IJ}\in{\mathbb Z},\, \det(\bf m)=\pm1 \right\}$. To explain why  this group   must indeed include all   invertible matrices with integral entries and determinant $\pm 1$, we need to recall that  each 2D  simple (Bravais) lattice   is  described by two linearly independent vectors $\lbrace {\bf e}_I\rbrace,\;I=1,2$, representing the  lattice basis.   In fact, infinite number of basis vectors exist describing the same lattice structure;   two bases  ${\bf e}^0_I$ and  ${\bar{\bf e}}_I$ describing the same lattice are related through 
${\bf {e}}^0_J  =  m_{IJ}   \bar{\bf e}_I$ with $ { m}_{IJ} \in {\mathbb Z}\,$ a unimodular matrix  with integer entries. Therefore, it is exactly the  symmetry group $GL(2,\mathbb{Z})$ which accounts for the lattice invariant   deformations. 

It can be shown   \cite{Pitteri1984-bp} that $GL(2,\mathbb{Z})$  constitutes the  finite strain extension  of the  crystallographic point group  $P(\mathbf{e}_I)$ which   describes  material symmetries in classical continuum elasticity \cite{Coleman1964-ry,Truesdell2004-ue}. 
 In the presence of $GL(2,\mathbb{Z})$  symmetry,  the space of metric tensors $\bC$ partitions into   periodicity domains, each one containing an  energy well equivalent to the reference one
\cite{Folkins1991-em,Parry1998-sv}.  Therefore, if  we know the structure of the energy $\phi$ in one of such domains,  we can use the $GL(2,\mathbb{Z})$ symmetry   to find its value in any other point of the  configurational  space of metric tensors $\bf C$  described by its  three significant components   $C_{11}, C_{22}$ and $C_{12}$. 
Since the $GL(2,\mathbb{Z})$ invariance  does not concern lattices with different volumes, it is sufficient to   focus on the 2D subspace of the configurational space  described by the condition $\det{\bf C}=C_{22}C_{11}-C^2_{12}=1$ \cite{Engel1986-tm,Conti2004-sv}.        
\begin{figure}[h!]
\includegraphics[width=1.0\columnwidth]{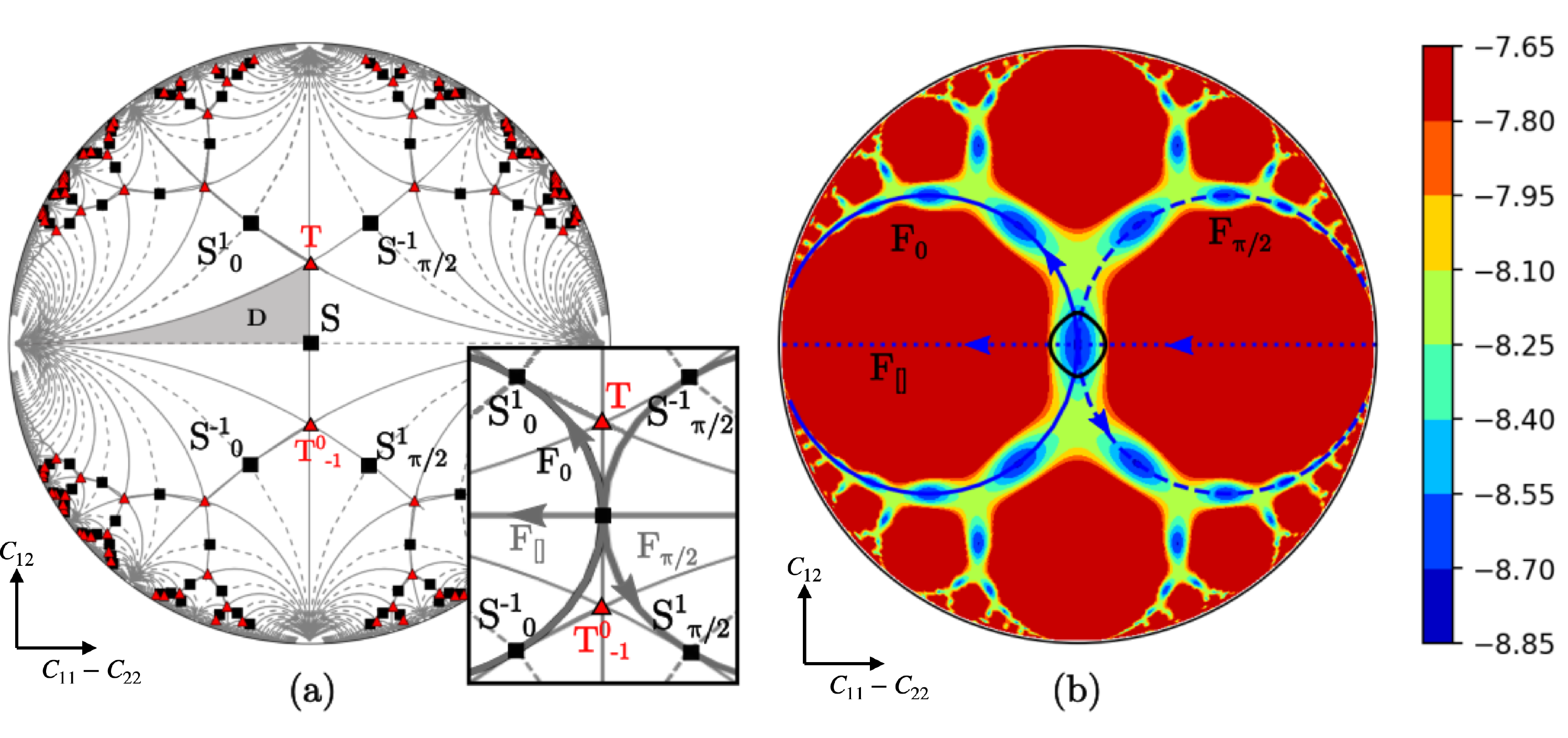}
\caption{\footnotesize{(a)  The configurational space of metric tensors with $\det{\bf C}=1$ (Poincaré disc). Infinite  families of equivalent  square and triangular lattices are shown  by squares ${\bf S}$ and triangles ${\bf T}$, respectively; $\bf D$ is the minimal periodicity domain; 
(b) The    energy landscape  obtained  using an    interatomic potential from \cite{Boyer1996-wt}; a cutoff was applied   to improve the visibility of the low-energy valleys. Blue circles: simple shears paths ${\bf F}(\alpha, 0 )$ (continuous) and ${\bf F}(\alpha,\pi/2)$ (dashed). Dotted straight line:  pure shear path  $\mathbf{F}_\rectangle(\alpha)$. The dark oval in (b)  represents the effective  yield surface.}
}\label{fig:atlas}
\end{figure}

   To visualize the implied  tensorial periodicity of the function $\phi(\bC)$,  it is convenient to stereo-graphically  project  the  hyperbolic surface  $\det{\bf C}=1$ on a 2D  disk of unit radius (Poincar\'e disk).
In this   mapping the  point $(x,y)$ on the disk represents the  point $(\widehat{x},\widehat{y},\widehat{z})=((C_{11}-C_{22})/ 2 ,C_{12},(C_{11}+C_{22})/2)$ on the hyperbolic surface  with  
$$
x=\frac{(\frac{C_{12}}{C_{22}})^{2}+(\frac{1}{C_{22}})^{2}-1}{(\frac{C_{12}}{C_{22}})^{2}+((\frac{1}{C_{22}})+1)^{2}},                      
y=\frac{2(\frac{C_{12}}{C_{22}})}{(\frac{C_{12}}{C_{22}})^{2}+((\frac{1}{C_{22}})+1)^{2}}.
$$
In these relations, once $C_{22}$  and $C_{12}$ are given, the component $C_{11}$ is determined by the condition $ C_{11}C_{22}-C_{12}^2=1$.

We illustrate the $GL(2,\mathbb{Z})$-induced  tessellation of the configurational surface  in Fig.\ref{fig:atlas}(a). The subdomain   $D = \{ 0<C_{11}\le C_{22},\quad 0\le C_{12}\le  C_{11}/2 \}$, highlighted in grey in Fig. \ref{fig:atlas}(a), contains metric tensors forming the   'minimal' periodicity domain on the surface $\det{\bf C}=1$.  It corresponds to the  'minimal' choice for the  lattice vectors $\tilde{\mathbf{e}}_1,\tilde{\mathbf{e}}_2$,  selected according to the algorithm \cite{Engel1986-tm}:
  (step 1) $\tilde{\mathbf{e}}_1$ is the shortest lattice vector;
(step 2) $\tilde{\mathbf{e}}_2$ is the shortest lattice vector not co-linear with $\tilde{\mathbf{e}}_1$ and for which  the sign is chosen in such a way that the angle between the two is acute. 
 The ensuing  basis is known as having the  'reduced form of Lagrange'. It is related to the original basis through the relation $\tilde{\mathbf{e}}_b=m_{ab}\mathbf{e}_a$  from which the matrix $\mathbf{m}$  is readily identified.  
 The  algorithm of   Lagrange reduction,  can be also recast directly in term of metric tensors $\bf C$:  
(step 1) if $C_{12}<0$, change sign to $C_{12}$; 
(step 2) if $C_{22}<C_{11}$, swap the two components; 
(step 3) if $2C_{12}>C_{11}$ set $C_{12}=C_{12}-C_{11}$ and $C_{22}=C_{22}+C_{11}-2C_{12}$. 
For a generic metric $\bf C$, such iterative scheme produces the  'minimal' equivalent metric  $\tilde {\bf  C} \in D$.  

Note that in Fig. \ref{fig:atlas}(a) we marked by small black squares the points corresponding   to the location of equivalent   lattice  configurations with square symmetry. The equivalent  lattice configuration with hexagonal symmetry (corresponding to triangular lattices) are marked by small red triangles. Simple  and centered rectangular   lattices  form one parametric families and are  marked by lines.  For the graph theory representation of the relation among equivalent metrics which also reveals additional  crystallographic aspects of the underlying energy wells structure,  we refer to  \cite{Denoual2016-qm,Gao2019-wa}. It is also instructive to compare the geometrically exact, finite strain, fully  tensorial periodicity of the configurational space described above,  with an approximate one,  generated by theories which rely on linearized strains, see for instance  \cite{Minami2007-ew,Onuki2003-ln,Carpio2005-bv,Geslin2014-ad}.

\begin{figure}[hbt!]
\centering
\includegraphics[scale=0.25]{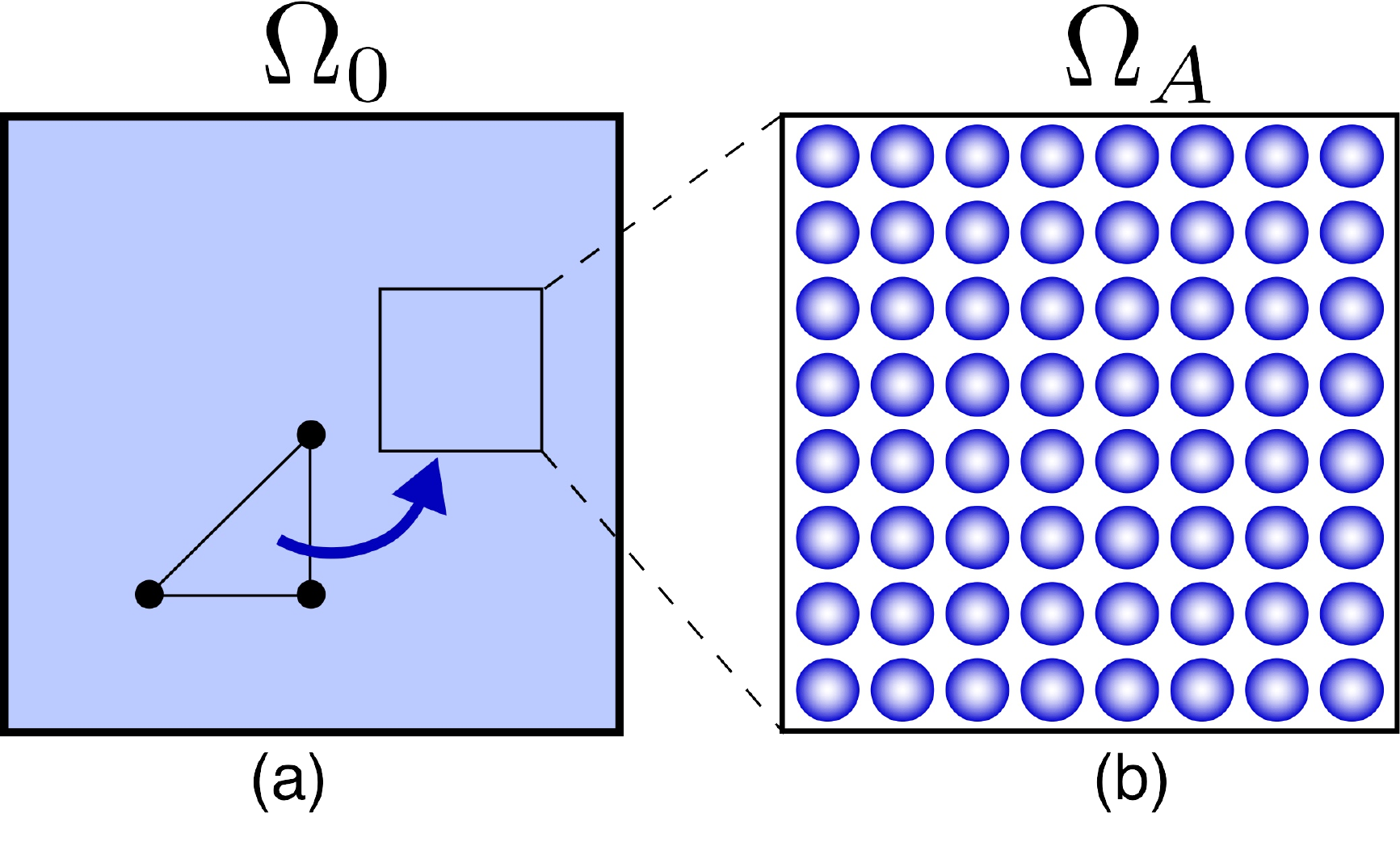}
\caption{\footnotesize (a)  Schematic representation of the  triangulated macroscopic  domain $\Omega_0$ where the mesoscopic Landau energy is defined on each of the triangles; (b)  the   atomistic domain $\Omega_A$ which represents the response of each of the macroscopic triangles and where the summation in \eqref{eqn:cg_sum} is performed; here the blue balls represent the actual atoms. }   \label{fig:fignew}
\end{figure}

 \emph{ Energy landscape.} The  multi-well (periodic)  Landau potential $\phi(\bC)$  can be computed   from a micro-scale theory using the Cauchy-Born rule  \cite{Ericksen2008-kx}. Suppose, for instance,   that the atomic interactions are pairwise and that the interatomic potential  $V({\bf r})= V(r)$,  where r is the distance between the atoms, is known. Suppose then  that material points in a representative volume $\Omega_A$ undergo an affine  deformation,  see Fig. \ref{fig:fignew}.
$Y_i(\boldsymbol{\boldsymbol{X}}) = X_i + u_i(\boldsymbol{X}),$
where   ${\bf X}$ and ${\bf Y}$ are the coordinates of atoms in the undeformed and deformed states, respectively, and   $u_i(\boldsymbol{X})$ is the displacement vector. The  vectors connecting two atoms in the reference configuration are $R_i  = X_i - X_i' $ and in the deformed configuration are $r_i  = Y_i - Y_i' $.  If the deformation is affine, we can write   $r_i  = F_{ij} R_j$ where 
the deformation gradient  is 
$F_{ij} =  \partial Y_i/\partial X_j = \delta_{ij} + \partial u_i/\partial X_j$. To compute the energy density $\phi( \bC) $   we need to   account for the   deformation of each atomic bond and then  average over the domain $\Omega_A$. Given that     in the $2D$ case one  can express the interatomic potential   in the form
$V(\sqrt{ R_1^2 C_{11} + 2 R_1 R_2 C_{12} + R_2^2 C_{22} })$,   we can   write
 \begin{equation}
\label{eqn:cg_sum}
\phi( \bC) = \frac{1}{2\Omega_A}\sum_{\boldsymbol{X}} \sum_{\boldsymbol{X'}\in\mathcal{N}(\boldsymbol{X})} V  \Bigl( \sqrt{R_i C_{ij} R_j} \Bigr),
\end{equation}
where   $C_{ij}=F_{ki}F_{kj}$ and the internal summations involves  all points $\boldsymbol{X'}$ belonging to the cutoff neighborhood $\mathcal{N}(\boldsymbol{X})$.

 The  periodic energy landscape   used in  subsequent numerical experiments, was constructed using the above algorithm and  based on a particular   interatomic potential   
\begin{equation}
\label{eq1:sm}
V(r) = -2/e^{-8(-1.425 + r)^2} -2/e^{-8(-1 + r)^2} + 2/r^{12} 
\end{equation}
constructed in \cite{Boyer1996-wt} to ensure that the ground state is a square lattice. In order to perform the sum in  \eqref{eqn:cg_sum}, we  used the  reference square lattice composed of $8\times8$ atoms with the lattice distance  $r_0=1.0658$. It is clear that with  representative volume $\Omega_A$  growing to infinity,
the function $\phi( \bC)$  develops  the  periodic structure consistent with the symmetry  of lattice invariant shears, e.g.  \cite{Juan1993-cu,Sunyk2003-pc}.  The choice of the cutoff scale, limiting the volume of volume $\Omega_A$,  is dictated in each problem  by the required range of  (almost) periodicity of the configurational energy landscape.  In fact, in view of the global  symmetry of the energy,  it is sufficient to construct  the potential $\phi({\bf C})$  using  \eqref{eqn:cg_sum} only  inside the minimal  periodicity domain  $D$, even if approximately,  using a finite rather than an infinite representative volume.   Then, as we have seen above, for an arbitrary metric tensor $\bf C$ one can produce  the appropriate symmetry transformation   $\bf m$ and  use the mapping $\tilde{\bf C}={\bf m}^T{\bf C} {\bf m}$   into $D$ to compute $\phi(\tilde {\bf C})=\phi({\bf C})$. The ensuing energy landscape is illustrated in Fig. \ref{fig:atlas}(b).


\begin{figure*}[hbt!]
\centering
\includegraphics[scale=0.15]{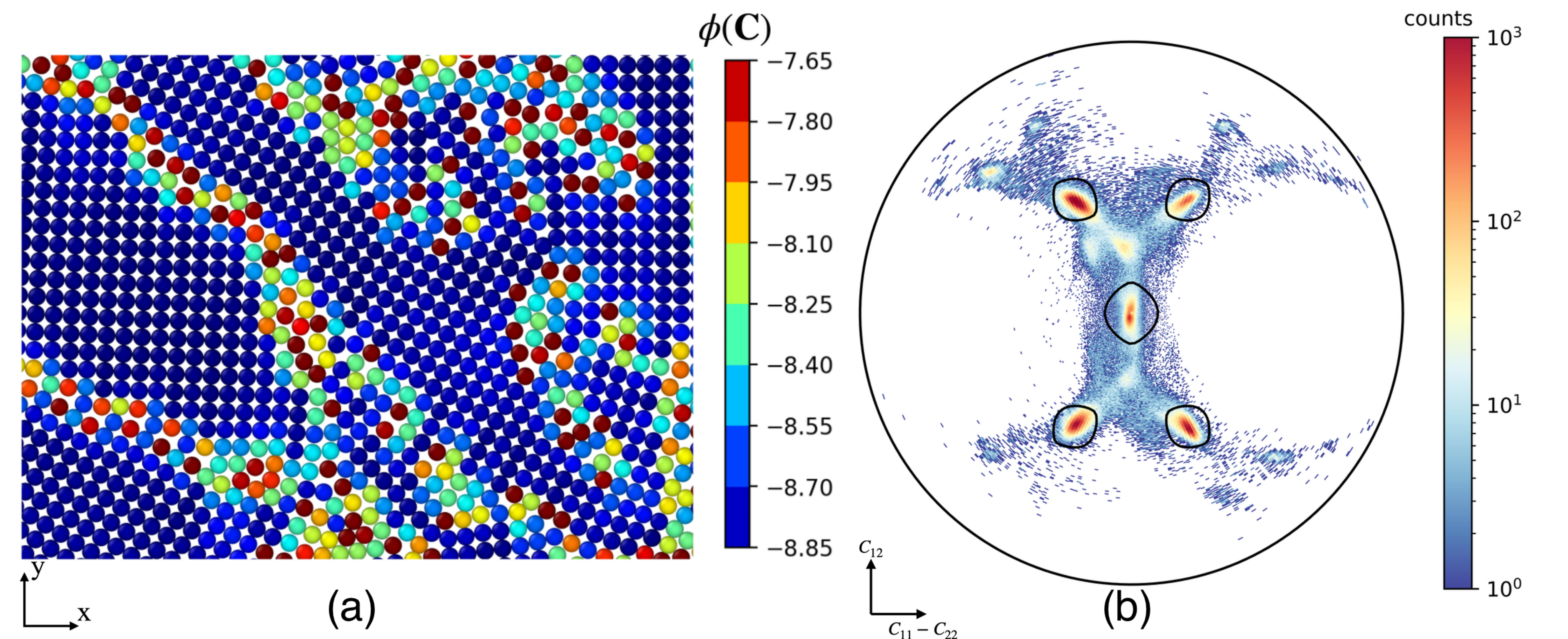}
\caption{\footnotesize{(a) A fragment of the post-avalanche pattern   in real space. The  balls represent mesoscopic finite elements; colors indicate the level of energy.
  (b) The  distribution of elements in the configurational space for the whole   post-avalanche pattern  obtained  by counting the number of elements inside individual configurational  bins. The dark ovals  in (b)  show  the effective  yield surfaces around the equivalent energy wells (pristine  crystal and its four  copies obtained by the smallest lattice invariant shears). 
 }} \label{fig:pvslj}
\end{figure*}

  Note first that  as a result of the imposed invariance, energy is periodic along simple shears of the type ${\bf F}(\alpha,\theta)={\bf I}+\alpha{\bf R}(\theta){\bf e}_1 \otimes{\bf R}(\theta){\bf e}_2$, where $ {\bf e}_ i$ is the orthonormal basis of the reference lattice, and ${\bf R}(\theta)$ is counter-clockwise rotation by the angle $\theta$, and $\alpha$ is the magnitude of the shear. In particular, conventional plastic mechanisms for a square lattice correspond to simple shears with  $\theta=0,\pi/2$.  Such  shears  are aligned with the two main close packed crystallographic (slip) planes and therefore correspond  to low energy valleys in the  configurational    landscape shown in Fig. \ref{fig:atlas}(b). 
  
  Note also that the loading paths ${\bf F}(\alpha,\theta)$  with  $\theta=0,\pi/2$ go through all  square wells.  In particular, the matrices $\textbf{F}({\alpha, 0})$ and  $\textbf{F}({\alpha, \pi/2 })$  with  integer  entries    mark the  bottoms of the equivalent energy wells and correspond to  lattice invariant shears. The energy barriers, separating   adjacent    wells   are relatively low which qualifies the corresponding tensorial directions as 'soft'.   On the Poincar{\'e} disk   such periodically modulated  'low energy valleys'  are described by  
circular trajectories, see  Fig. \ref{fig:atlas}(b).

%
\emph{Geometrical linearization.} Here it is appropriate to remark that the geometrically non-linear focus of MTM     is not redundant and instead has a crucial effect on the outcome of the numerical experiments vis a vis the results obtained in   geometrically  linearized models  \cite{Minami2007-ew,Onuki2003-ln,Carpio2005-bv,Geslin2014-ad}.  Indeed, consider again the two symmetry related shears  describing our 'soft' loading directions $\textbf{F}(\alpha,0)$ and $\textbf{F}(\alpha,\pi/2)$.
%
%
 In a geometrically linearized  description, the components of the infinitesimal strain tensor ${\boldmath \epsilon}=\frac{1}{2}\left(\nabla{\bf u}^T+\nabla{\bf u} \right)$ are the same for both paths:
\begin{equation}
\label{eqn:Elin}
{\boldmath \epsilon}=\frac{1}{2}\left[\begin{array}{cc}
0 & \alpha\\                     
\alpha & 0
\end{array}\right].
\end{equation}
Instead, in the geometrically nonlinear theory these two configuratiuonal directions are energetically equivalent but different. For instance,   the associated nonlinear strains  tensors  ${\bf E}=\frac{1}{2}(\nabla{\bf u}^T+\nabla{\bf u}+\nabla{\bf u}^T\nabla{\bf u})= (1/2)({\bf C}-{\bf I})$ are different at the second order in $ \alpha$ :
\begin{equation}
\label{eqn:Ezernov}
\textbf{E}_{0}=\frac{1}{2}\left[\begin{array}{cc}
0 & \alpha\\                     
\alpha & \alpha^{2}
\end{array}\right]\qquad \textbf{E}_{\pi/2}=\frac{1}{2}\left[\begin{array}{cc}
\alpha^{2} & \alpha\\                     
\alpha & 0
\end{array}\right].
\end{equation}

Note also that the  nonlinear terms ($\propto\alpha^2$) appear  along  the diagonal entries $E_{11}$ and $E_{22}$, which describe  the stretch along the initial lattice vectors ${\bf e}_{1}$ and ${\bf e}_{2}$. 
In the small strain limit these vectors do not stretch  and  the   constant volume requirement  reduces to the condition  that the trace of the linearized strain  is zero. Therefore, the linearized states of strain are located on the configurational plane  $tr({\bf C})=2$,  which is tangent to the configurational hyperboloid of the geometrically nonlinear theory $\det{\bold C}=1$ and therefore agrees with it only locally.
On this plane the two paths  with $\theta=0,\pi/2$, corresponding physically to the activation of two different slip systems,  merge  into one  and    the two geometrically distinct  lattice configurations $ \mathbf{F}(\alpha=1,\theta=0)$ and $ \mathbf{F}(\alpha=-1,\theta=\pi/2)$ are represented by a single   point. This causes a well known degeneracy  in the geometrically linear theory where one cannot distinguish between the two different slip systems and have to make additional phenomenological  assumptions to resolve the degeneracy \cite{Marano2019-he}.
 

\section{Numerical experiments}
 \label{sec:num}

In our numerical experiments a pristine crystal was represented by a homogeneous square domain $\Omega_0$  divided (triangularized) into $600 \times 600\times2$ elements  aligned with coordinate axes.  This sample was loaded quasi-statically in a hard device by applying on the boundary an incremental affine deformation $\mathbf{F}$ parametrized by a scalar parameter  $\alpha $. The response of the system is represented by the deformation of the elements described by the functions $\by(\bx; \alpha)$.

\emph{Loading path. }To ensure that the  response is generic,   
  instead of the  'soft' loading directions given by $\mathbf{F}(\alpha,\theta)$  with $\theta=0,\pi/2$, we used 
 the 'hard' loading  path corresponding to  pure shear  
\begin{equation}
\label{1}
\mathbf{F}_\rectangle (\alpha)
=
\left[\begin{array}{cc}
\cosh(\frac{\alpha}{2})-\sinh(\frac{\alpha}{2}) & 0\\
0 & \cosh(\frac{\alpha}{2})+\sinh(\frac{\alpha}{2})
\end{array}\right].
\end{equation}
The   mapping  \eqref{1} transforms squares into  rectangles and  we use here a  natural parametrization 
  $\alpha=2\log(\lambda)$,  where   $(\lambda,\lambda^{-1})$ are the two principal stretches \cite{Destrade2012-kc}; the corresponding path in the configurational space is shown by the dashed line in Fig. \ref{fig:atlas}(b).  At each step of such loading, parametrized by $\alpha$,  we perform  incremental energy minimization.   Effectively, we use continuation method and search  for a one parametric family of   equilibrium configurations. 
  
\emph{ Numerical method.}  We reiterate that in the numerical code, the reference body  $\Omega_0$ was divided into triangular finite elements. Therefore we traced the response $\by(\bx; \alpha)$  through    the  deformation of the   2D network of discrete nodes.  The   nodes $\bx$  can be  identified using integer-valued coordinates $ij$ and with each   node we associated a  cell defined by the basis vectors $\be_a(\bx)$, where $a=1,2$.  We then treated each triangular cell as a  finite element with linear  shape functions  $N_{ij}({\bf x})$. This allowed us to write the discrete displacement field in the form ${\bf u} ({ \bf x})={\bf u}_{ij} N_{ij} ({\bf x})$, where ${\bf u}_{ij} $ denote the values of displacement at node $ij$.   The discrete deformation gradient is then  $\bna\by= \mathbf{1}  + {\bf u}_{ij} \otimes \nabla N_{ij} $. 
  
 We recall that the elastic energy of a  cell $\phi( \bC)$ associated with node $\bx$,  is a function of the metric tensor $\bC=\bna\by^T\bna\by$.  To minimize the energy functional $W=\int_{\Omega_0} \phi d\Omega_0$  we  used a variant of conjugate gradient optimization known as the  L-BFGS algorithm~\cite{Bochkanov2013-lk}.  It finds  solutions of  the equilibrium equations 
$
\partial W/\partial {\bf u}_{ij} =\int_{\Omega_0} {\bf P}\nabla N_{ij}d\Omega_0=0,
$
where  ${\bf P}= \partial \phi / \partial\bna\by$,  that are reachable through algorithmically defined, effectively overdamped dynamics.   In view of the  hard device loading,   the positions of surface nodes were set to satisfy  ${\bf y}  =\mathbf{F}_\rectangle (\alpha) {\bf x}$.  

The simulations were performed on a fixed computational grid  and we did not seek any grid refinement because, due to the particular structure of the non-convexity of $\phi$, our solid system  degenerates and behaves mechanically as a liquid in continuum limit \cite{Fonseca1987-pd}. To avoid this unphysical behavior, we interpreted the mesh/grid size $h \sim N^{-1}$ as a regularizing (cut-off)  parameter of physical nature. 

As we have already mentioned, in physical terms   this parameter is the mesoscopic length at which the deformation can be considered homogeneous.  It is also a length scale at which  the ab initio Cauchy-Born energy of a  homogeneously deforming elastic element can be considered periodic in the range of strain which captures all the   relevant   energy wells.

%
%
  \begin{figure}[h!]
\centering
\includegraphics[width=.7\columnwidth]{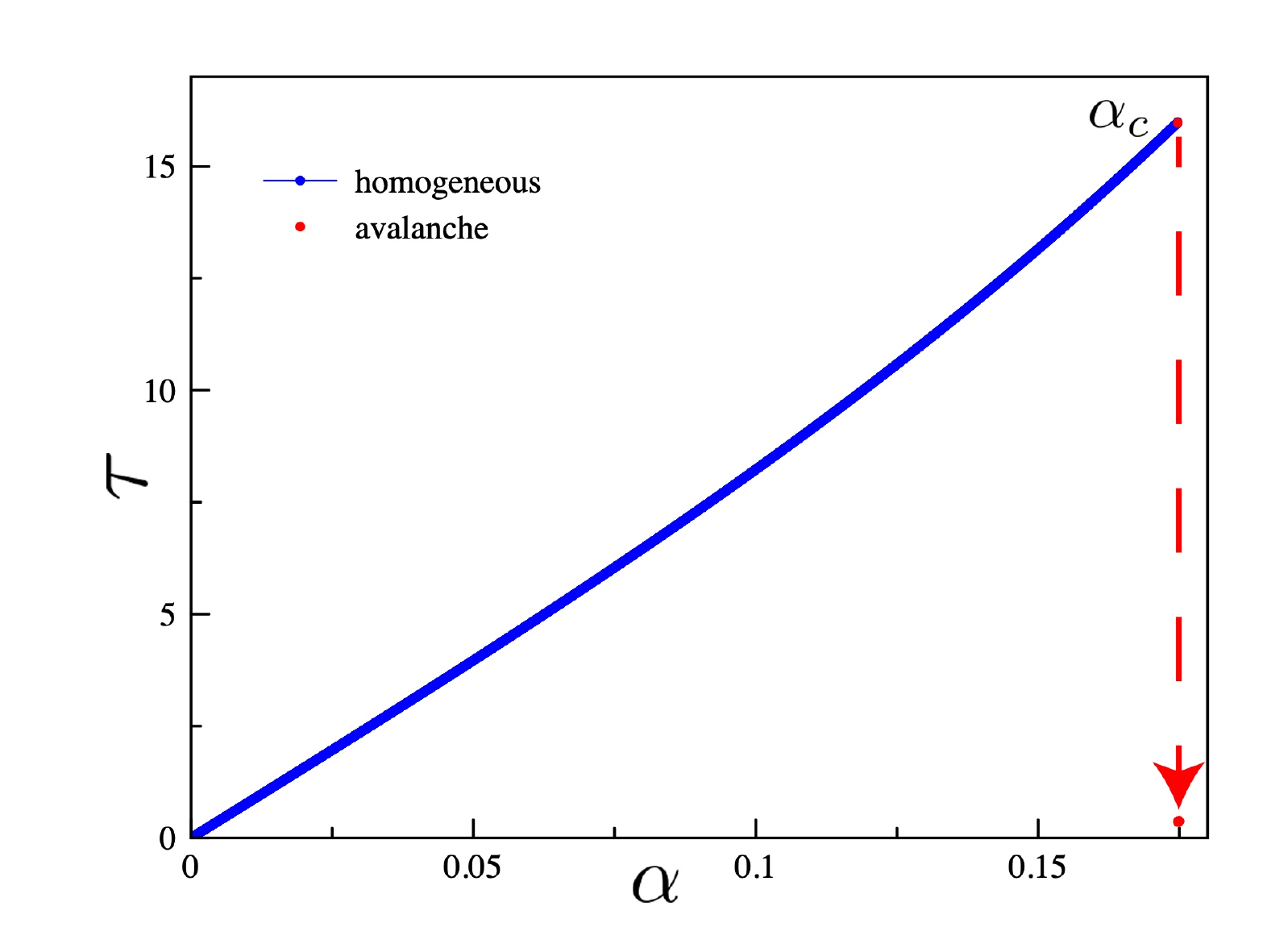}
\caption{\footnotesize{Quasi-static stress-strain response along the   path $\mathbf{F}_\rectangle (\alpha)$. The stress is projected on the loading direction and averaged over the sample: $\boldsymbol\tau(\alpha)  = \int_{\Omega_0}\mathbf{P}:(d\mathbf{F}_\rectangle (\alpha)/d\alpha)d\Omega_0$. The initial nonlinear elastic branch (in blue) corresponds  to  affine response. The stress drop (in red)  corresponds to system size dislocation avalanche at $\alpha=\alpha_c$. 
}} \label{fig:ssrect}
\end{figure}

\emph{Stress-strain response. } In Fig. \ref{fig:ssrect}, we present the  numerically generated quasi-static stress-strain response of a perfect crystal  loaded in a hard device along the  path $\mathbf{F}_\rectangle (\alpha)$.   We observe that first  the crystal deforms elastically as the  affine  mapping  $\mathbf{F}_\rectangle (\alpha)$   simply transforms squares into  rectangles. The homogeneous  elastic    branch  becomes unstable at  $\alpha=\alpha_c$  when  the   branch switching event (dislocation avalanche) takes place.  It is a system size collective dislocation nucleation forcing  the stress to drop precipitously to an almost zero level. Such an extreme relaxation of the  accumulated elastic stress through a catastrophic  avalanche suggests that  most of the nucleated dislocations end up topologically compensated  and therefore effectively screened. The ensuing  defect microstructure  is then  mostly composed of 'statistically stored'   (rather than 'geometrically necessary')   dislocations \cite{Arsenlis1999-nf,Rezvanian2007-jc,Nguyen2021-sp}.

\emph{Effective yield surface. } The   value of the critical  loading parameter $\alpha_c$  can be anticipated using  
the macroscopic Legendre-Hadamard (strong ellipticity) criterion \cite{Rice1976-ta,Ogden1997-rf,Bigoni2012-by, Van_Vliet2003-yg,Li2004-vb,Zhong2008-pj}
  \begin{equation}
\label{eqn:acou}
\mathcal{Q}_{ik}(N)=N_{J}{\mathcal A}_{iJkL}N_{L}>0\,,
\end{equation}
where the acoustic tensor $\mathcal{Q}_{ik}$,   is given by $\mathcal{Q}_{ik}=N_{J}{\mathcal A}_{iJkL}N_{L}$, while $\mathbf{N}$ and $\mathbf{m}$ are two arbitrary vectors defined in the Lagrangian (un-deformed) and Eulerian (deformed) configurations, respectively. Here ${\mathcal A}_{iJkL}$ is the tensor of elastic moduli:  
\begin{equation}
\mathcal{A}_{iJkL}=\frac{\partial^2\phi}{\partial F_{iJ}\partial F_{kL}}\,.
\label{eq:pkmod}
\end{equation}
A rigorous  analysis shows that the weak local stability of a homogeneous equilibrium in a hard device  is lost when the inequality (\ref{eqn:acou}) is no longer strict \cite{Grabovsky2013-ol}.
At the critical value of the loading parameter the  equality emerges in \eqref{eqn:acou} for some non-trivial $\bf N$ and $\bf m$. These two vectors represent,  respectively, the orientation and the polarization of the incipient unstable mode. 

Note that the instability  condition (\ref{eqn:acou})  can be also fully projected into the  Eulerian space. To this end we need to introduce  the Eulerian moduli 
$
\texttt{a}_{ijkl}=F_{jR}F_{lS}{\mathcal A}_{iRkS}.
$
The  Eulerian version of  the acoustic tensor can be then written in the form 
$
  q_{ik} (n) =n_{j}n_{k}\texttt{a}_{ijkl},
$
which produces  the stability criterion:
 \begin{equation}
\label{eqn:adet_q}
\det (\bf q({ n}))>0\,.
\end{equation} 
Applying the  condition $\det  \bf q =0$ to our loading  path $\mathbf{F}_\rectangle (\alpha)$ we   verified  that 
the numerically obtained value of the instability  threshold    
   $\alpha_c\approx0.176$   is in excellent agreement with the  theoretical prediction.
 
 We similarly  applied the condition (\ref{eqn:adet_q}) to a large family of   loading trajectories. By interpolating the resulting instability (spinodal) thresholds, we reconstructed the boundary of the affine response, which is  illustrated in black in Fig.~\ref{fig:atlas}(b). It can be viewed as representing an apparent  "yield surface" since an ideal crystal loaded in a hard device will deform homogeneously following the imposed loading path as long as the loading trajectory is contained within such stability region.   Note, however, that in a generically loaded inhomogeneous state,  the   \emph{interacting} configurational elements can   become unstable in a   broad area adjacent to this sharp  stability boundary, as it is evidenced, for instance,  by the studies of dislocation nucleation during highly heterogeneous micro-indentation tests \cite{Miller2008-jk,Garg2016-kz}.  
 

\emph{Catastrophic avalanche.} We now turn to the description of the system-size instability experienced by a perfect, defectless crystal at $\alpha=\alpha_c$.  As we have already mentioned, the breakdown of the affine (elastic) state $\bf y=\mathbf{F}_\rectangle (\alpha) \bf x$  takes the form of an abrupt drop of both stress and energy, apparently  signaling an almost pristine-to pristine transition. In reality, the instability causes the originally homogeneous crystal to rearrange its configuration between the neighboring equivalent energy wells which  proceeds through collective  generation and self-organization of mutually neutralized  dislocations of both signs.  

We observe that, as long as the  affine configuration remains  stable, the  points, representing different elastic elements   in the configurational space are all superimposed  (have the same value of  $ \bf C $) and follow together the prescribed loading path.  After the instability, the configurational points spread around the configuration space with several equivalent energy  wells getting  populated. During such configurational spreading   the crystal is deforming plastically as  the elastic  energy is slipping away irreversibly. It is assumed to be lost (dissipated) by either mechanical radiation or through thermalization and eventual heat conduction. 

\begin{figure}[h!]
\centering
\includegraphics[width=.8\columnwidth]{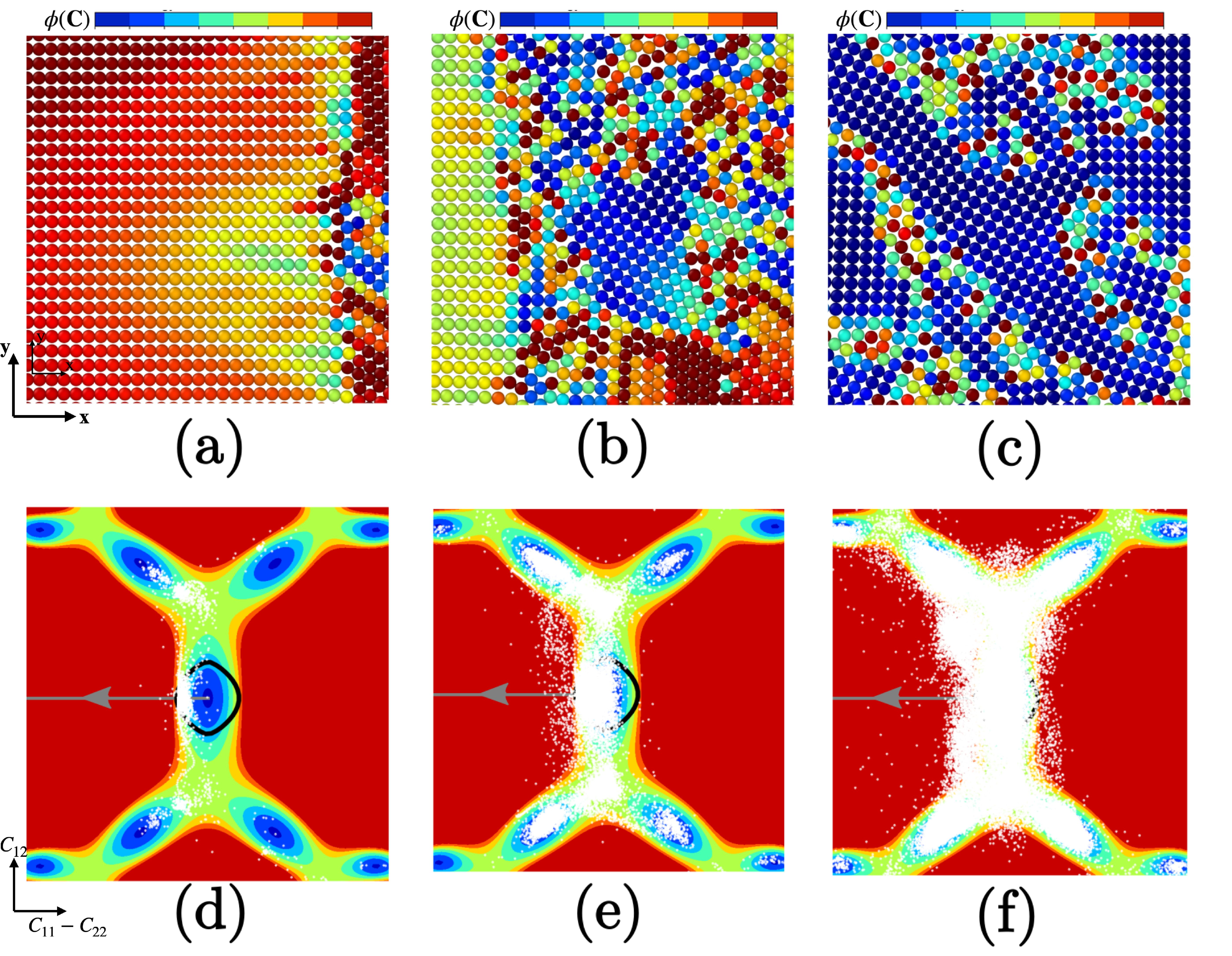}
\caption{\footnotesize{The unfolding of a dislocation avalanche: (a,b,c) in real space (colors indicate the level of strain energy density), (d,e,f) in the configurational space of metric tensors. 
   Large red/yellow region in (a) corresponds to elastically stressed homogeneous rectangular  configuration of the original lattice. Large blue cells in (c) correspond to symmetry related versions of unstressed square lattices. 
White dots in (d,e,f)  show the progressive spreading of the configurational points representing individual elastic elements. The dark ovals show the effective yield surfaces around the   energy well  corresponding to pristine crystal  
}} \label{fig:pvslj}
\end{figure}

In the physical space, as a result of  such  catastrophic system-size avalanche, the homogeneously deformed lattice is   replaced by a  complex texture of variously oriented patches of the unstressed lattice.  In Fig. \ref{fig:pvslj}(a,b,c)  we show the snapshots  of the evolving element lattice  during  the  avalanche  using  fast numerical time.  
\begin{figure*}[hbt!]
\centering
\includegraphics[width= 1.6  \columnwidth]{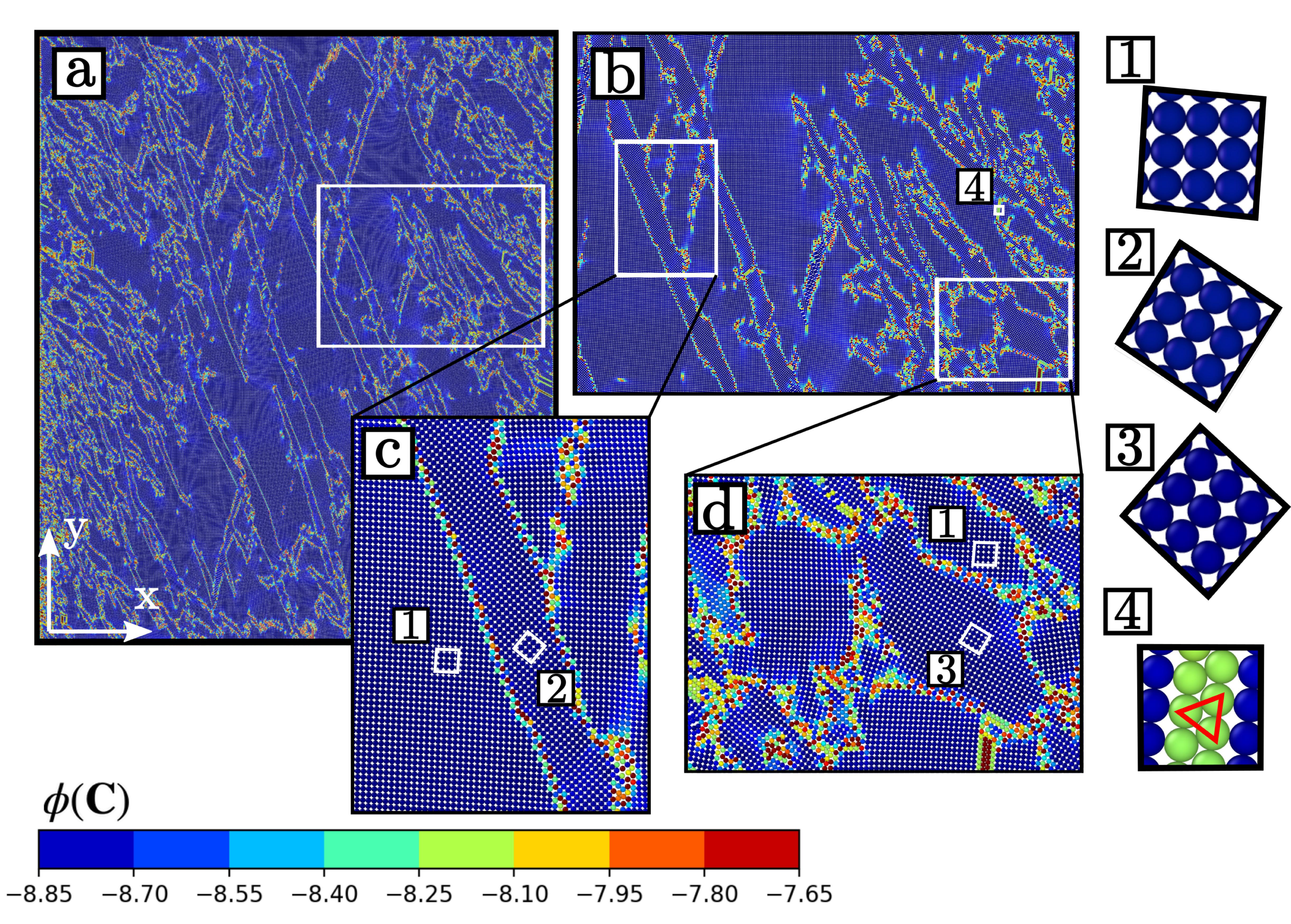}
\caption{\footnotesize{Post-avalanche pattern shown  at different scales. Colors indicate the level of energy.
 Insets (1-3):   relative rotation of unstressed square patches.  Inset (4):  a fragment of a metastable   triangular  lattice.  Here N=600. 
}} \label{fig:patt}
\end{figure*}  

 The spreading of the corresponding cloud in configurational  space  (Poincaré disk) is illustrated in Fig. \ref{fig:pvslj}(d,e,f). 
In particular,  Fig.~\ref{fig:pvslj}(f) shows the   post-avalanche  distribution of the deformed elements.  One can see that after the instability, most of the  homogeneously stretched (rectangular)    elements snap back to the reference energy well $\bf S$. However,  a significant percentage of elements also spreads right away over  the two symmetric energy wells ${\bf S}^0_1={\bf F}( 1, 0)$ and   ${\bf S}^0_{-1}={\bf F}( -1, 0)$,   which corresponds to the activation of single slip plasticity. The other two symmetric energy wells  ${\bf S}^{\pi/2}_1={\bf F}( 1, \pi/2)$ and   ${\bf S}^{\pi/2}_{-1}={\bf F}( -1, \pi/2)$   get populated eventually, but only at the end of the avalanche. This indicates that the second main slip system is  activated as well. Finally, some elements appear to be locked in the shallow local minima $\bf T$ and $\bf T^0_{-1}$ describing the triangular lattice with hexagonal symmetry; such elements appear mostly as the components of Shockley-type partials \cite{Medlin2019-je}. The limited occupation by  the elastic elements of spaces  outside  the  energy wells  reflects the presence of dislocation cores and, more generally,  highly heterogeneous defect structures.  
 
We illustrate the   gradual unfolding of the system size avalanche in our two movies (presented in the Supplementary Material),  which  reveal  the transient 'computational dynamics' hidden behind the apparently discontinuous avalanche. In particular, our \textbf{Movie S1} shows the fast-time evolution of the field ${\bf y} ({\bf x})$ (coloring shows the level of the non-diagonal component of  Cauchy stress, red-high, blue-low).  The complementary \textbf{Movie S2}  presents the same fast-time evolution but now of the non-affine component of the  field  ${\bf u}({\bf x}) = {\bf y}({\bf x}) - {\bf F_\rectangle}(\alpha_c){\bf x}$ represented by   blue arrows.
  
As we see in these movies,  the massive dislocation nucleation starts heterogeneously at the vertical boundaries of the body which play the role of effective lattice defects. It first proceeds in the form of two slowly propagating fronts,   originating on these boundaries and separating the remaining affine from  the growing non-affine configuration.  Eventually, these fronts get destabilized by the separate non-affinity originating on the horizontal parts of the boundary which leads to the fast development of considerable spatial complexity. The avalanche ends with another slow stage where the emerging pattern goes through a maturation stage. 

Note that if we interpret our  'computational dynamics' as a flow of the material, we may say that as the avalanche unfolds, a relatively 'laminar' flow is replaced by distinctly 'turbulent'  vortex dynamics. However, this spur of activity   is only transient and eventually its  intensity   decays, leaving behind a complex and manifestly multi-scale spatial pattern. Similar,   correlated non-affine transient  fluctuations during plastic avalanches have been extensively documented in (quasistatic) experiments on granular solids and even interpreted as 'granulence', see for instance  \cite{Misra1997-cy,Radjai2002-cj,Combe2013-zf,Oyama2019-ub,Sun2022-vt}. Turbulent-like 
displacement fields have been also found to accompany plastic avalanches in amorphous materials. The underlying manifestly non-Gaussian  fluctuations have been sometimes associated with   a correlated endogenous   “noise” ,  e.g. \cite{Goldenberg2007-ao,Ruscher2019-sc,Cui2022-cx}. We discuss the appropriateness of the  use of such   metaphors for the description of   plastic avalanches, observed in our numerical experiments,  in Section  \ref{turb}.

\begin{figure}[hbt!]
\centering
\includegraphics[width= 1. \columnwidth]{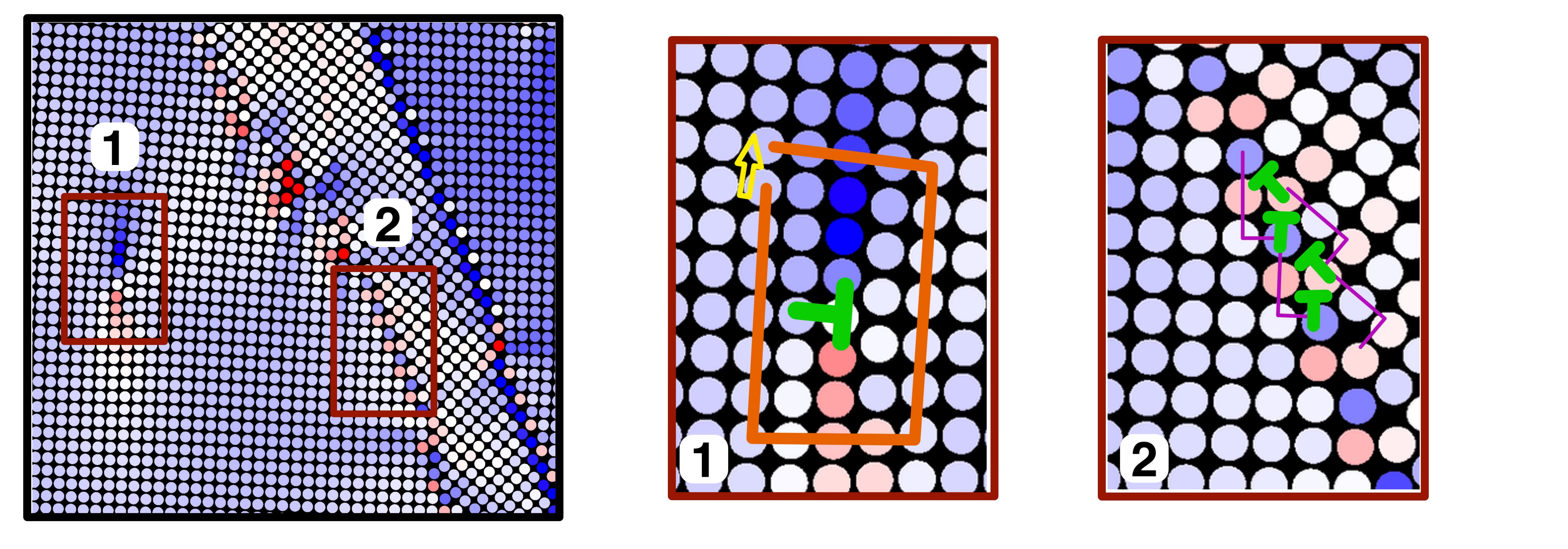}
\caption{\footnotesize{A fragment of the post-avalanche pattern  in physical space extracted from Fig. \ref{fig:patt}(a); colors indicate the level of the non-diagonal component of Cauchy stress.  The zoom into the fragment (1) shows a  single dislocation inside a rotated grain;  the open circuit around it   indicates   the corresponding Burgers vector    is non-zero with the length equal to  the lattice spacing, see the yellow arrow.  The zoom into   the fragment (2) shows the    boundary between the rotated patches  whose dislocational structure  reveals  the  $\Sigma 5$ type grain boundary.   }
}  \label{fig:dislo}
\end{figure}

\emph{Post avalanche pattern. }The detailed post-avalanche pattern in the physical space is shown in Fig.~\ref{fig:patt}(a).   Upon magnification we see a  complex arrangement of apparently randomly rotated unstressed square lattice patches circumscribed by energy carrying  boundaries, see Fig.~\ref{fig:patt}(b-d).    A salient feature of this  pattern is the ubiquitous presence of the lattice patches rotated at $\pi/2$ with respect to the undeformed reference state, see more about this below. We also see the ubiquitous presence of the  fragments of triangular lattice, like the one shown  in the inset in Fig.~\ref{fig:patt}, which serve as elements of the stacking fault-type interfaces and also contribute to the structure of at least some  dislocation cores.  In fact, a detailed study of the   deformation of individual elements  reveals the presence of both,  isolated    dislocations  and  the   dislocational-rich extended  lattice defects  where dislocations interact strongly and their cores may be distorted,  see  Fig. \ref{fig:dislo} (1,2).   In particular,  the  observed high-energy   boundaries which  separate the extended rotated patches,  can be viewed as   composed of   interacting dislocations   locked in stable, wall-type configurations, see  Fig. \ref{fig:dislo}(2). In general, the identification of  particular lattice defects and extracting their core structures in mesoscopic computations, where we  operate with atomically  blurred images, is an even more   challenging task  than  in the case of molecular dynamics \cite{Stukowski2012-gp,Elsey2014-yh,Barrett2017-xc,Bertin2020-hv}. In this sense our  reference to particular lattice defects emerging in   mesoscopic simulations can be viewed only as suggestive.

\emph{Molecular statics.} The natural question is then: how realistic is the observed picture? To corroborate  the  predictions of the MTM-base numerical experiments, we   performed a set of parallel molecular statics (MS) simulations employing the same interatomic potential \eqref{eq1:sm}. The role of elastic elements in such simulations was played by individual atoms, in other words,   we identified for simplicity our regularization length scale with interatomic distance. It is clear that in this case only qualitative agreement between the micro-and meso-scopic  pictures can be expected.

In our  MS numerical experiments the positions of the atoms for the given boundary conditions were determined by minimizing the   energy of a system composed of $N_A$ atoms.  It can be written in the form $\Pi = \frac{1}{2	}\sum_{\alpha}^{N_A} \sum_{\beta,\beta\neq \alpha}^{N_A} V  (r^{\alpha\beta})$, where  $r^{\alpha\beta}$ is the distance between the atoms   $\alpha$ and $\beta$.
As we have already mentioned, the interatomic potential  $V(r^{\alpha\beta})$   was taken from \eqref{eq1:sm}.   The positions of the atoms were found by solving the equilibrium equations  $ d\Pi/d {\bf r}^\lambda = 0$, where  $\lambda=1,\dots, N_A$.  To solve these  equations we used the L-BFGS algorithm~\cite{Bochkanov2013-lk} which builds a positive definite linear approximation of these equations allowing one to make a quasi-Newton step lowering the energy $\Pi$.   To impose the hard device type boundary conditions we applied affine displacements (of a pure shear type $\mathbf{F}_\rectangle (\alpha)$)  to the atoms within the boundary layer of a small thickness.  The amplitude of the loading was incrementally increased and kept fixed during each relaxation step. 

\begin{figure}[h!]
\includegraphics[width=.99\columnwidth]{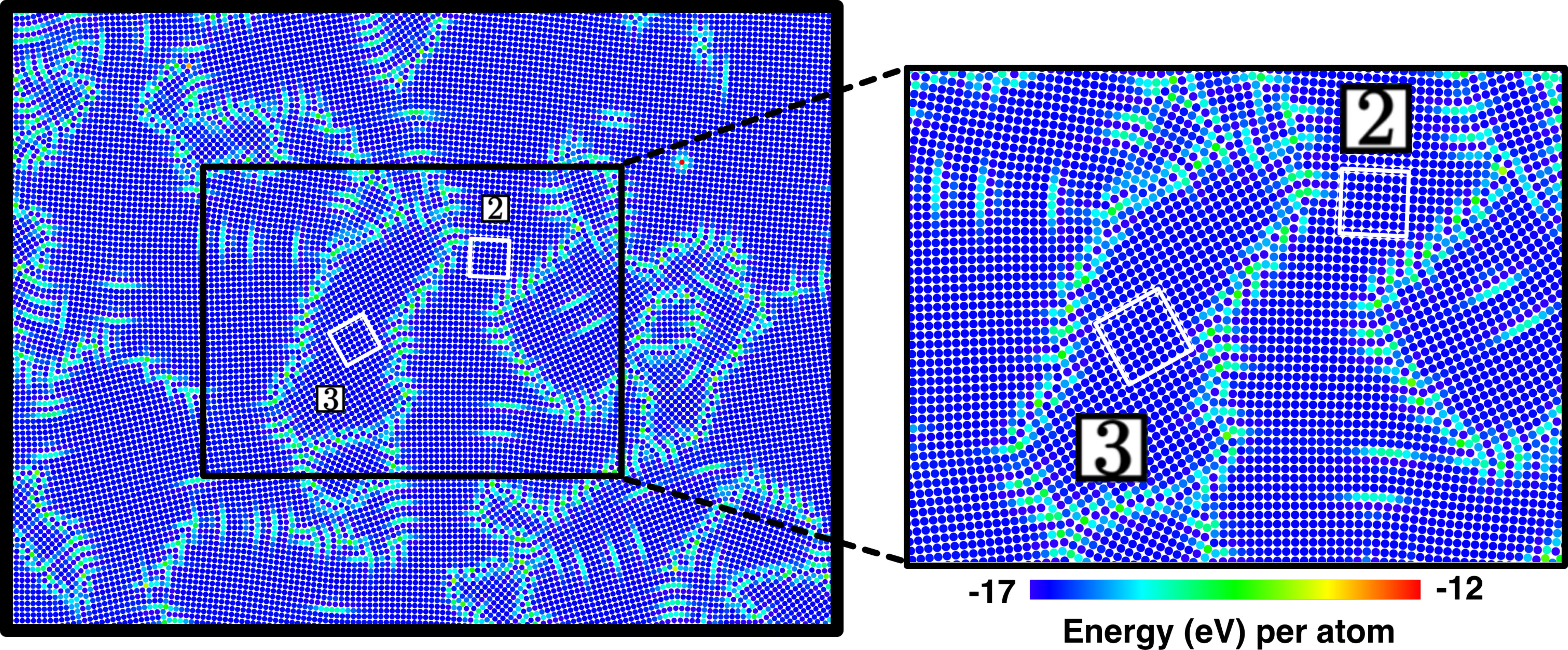}
\caption{\footnotesize{Final positions of the atoms  in our parallel MS study of a system size  plastic avalanche showing  a  fragment of the post-avalanche pattern.  Atoms are colored according to the level of their  potential energy: red-high, blue-low. The magnification on the right shows  an example of  an inelastically  rotated (micro-twinned) domain; numbers have the same meaning as in Fig. \ref{fig:patt}.
 Such  low-energy domains  are separated by high-energy dislocation-rich grain boundaries.      }
}\label{fig:sm1}
\end{figure}

As in MTM,  in our   MS  tests the pristine crystal was deforming homogeneously till the  critical value of the loading parameter $\alpha$  was reached. At the critical  level of strain, which is close to the theoretical  prediction,   the system size   plastic avalanche took place. In Fig. \ref{fig:sm1}, we present a fragment of the    post avalanche  configuration. It  shows the anticipated grain structure. One can see that the misoriented low energy  patches of original lattice  are practically unloaded. The energy is again localized on the highly dislocated  inter-grain boundaries. The inset shows a magnified version of the   $\pi/2$  rotation of one of the patches   relative to  the orientation of the original square lattice. Other differently oriented patches are visible as well forming collectively a complex crystallographic texture. The overall picture is   similar to the one obtained in our MTM-based experiments   which corroborates its basic conclusions. In this paper  we do not perform quantitative comparison that would require, in particular,  the discussion of the delicate role of the internal scales in  the two models; the corresponding analysis will be presented in a separate paper.

\section{Inelastic rotations}
\label{sec:ie}
In the previous section, we have seen that the MTM approach allows one to trace the emergence of complex multi-grain textures and study the formation of the supporting dislocation patterns. In this section, we show that MTM  also offers a suitable framework for a deeper theoretical understanding of the observed misorientations angles between the ensuing grains.
 
\begin{figure}[h!]
\centering
\includegraphics[width=.65\columnwidth]{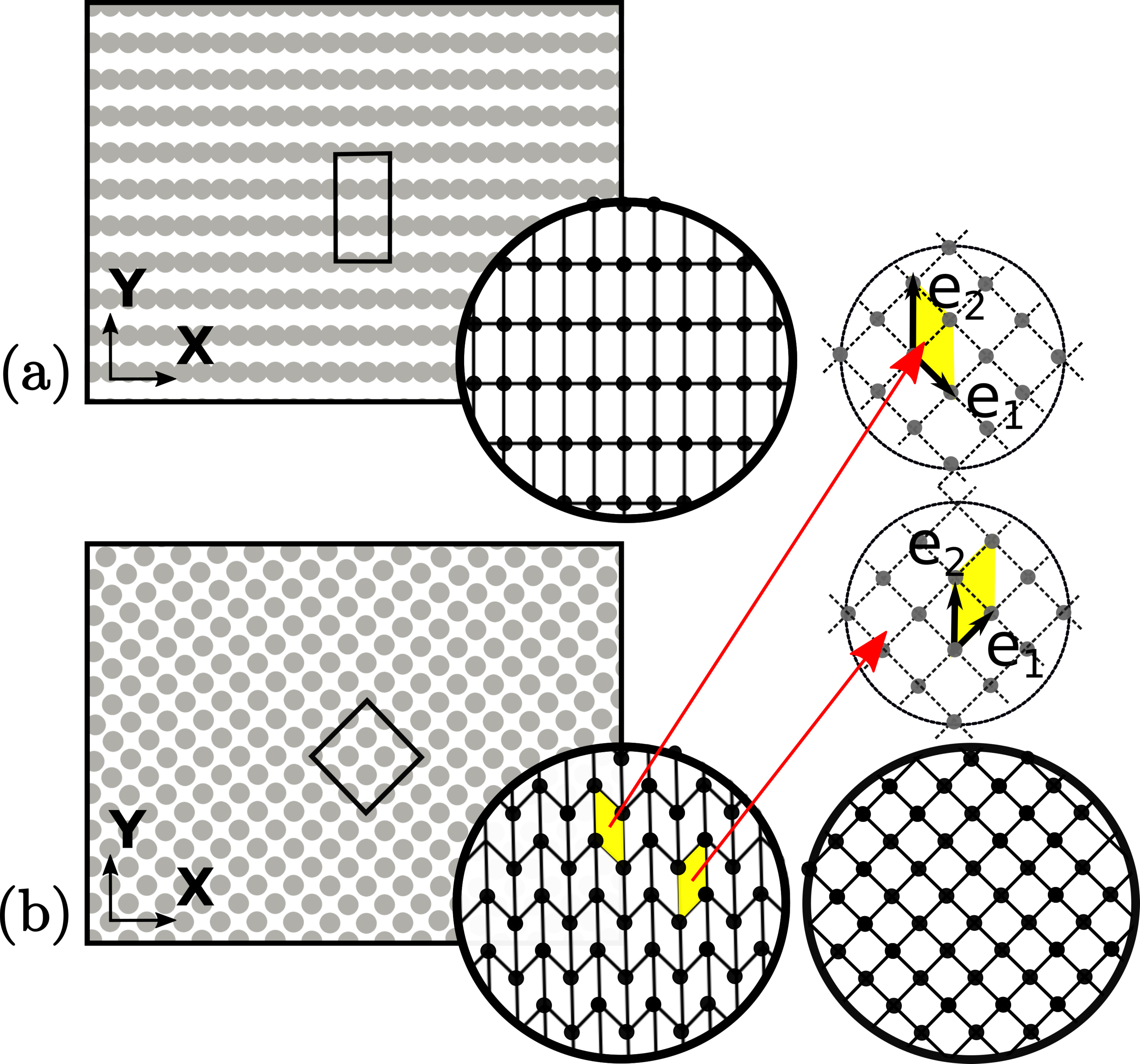}
\caption{\footnotesize {A detail of the crystal structure deformed along the rectangular path,  before and   after the  instability. The apparently rotated square lattice is obtained by a combination of  microscopically compatible sheared configurations. } }
\label{fig:skinny_patt2}
\end{figure}

The microscopic nature of large rotations in Fig.~\ref{fig:patt} can be understood if we follow the deformation of the individual elements.  Thus, the elemental triangulation  of the coexisting patches of type 1-3 in Fig.~\ref{fig:patt}  reveals that an apparent rigid rotation at the macroscale is, in fact,  a disguised micro-twin mixture of the elements of the types ${\bf R}(\pi/2){\bf S}^0_1$ and ${\bf S}^0_{-1}$. The zoom on the two configurations, before and  after the instability, is shown in Fig. \ref{fig:skinny_patt2}.  It affirms that inside the rotated patch the (loaded) rectangular lattice is transformed into the (unloaded) square lattice which turns out to be a crystallographically exact mixture of microscopically compatible sheared square lattice configurations.   

\begin{figure*}[hbt!]
\centering
\includegraphics[width=1.5\columnwidth]{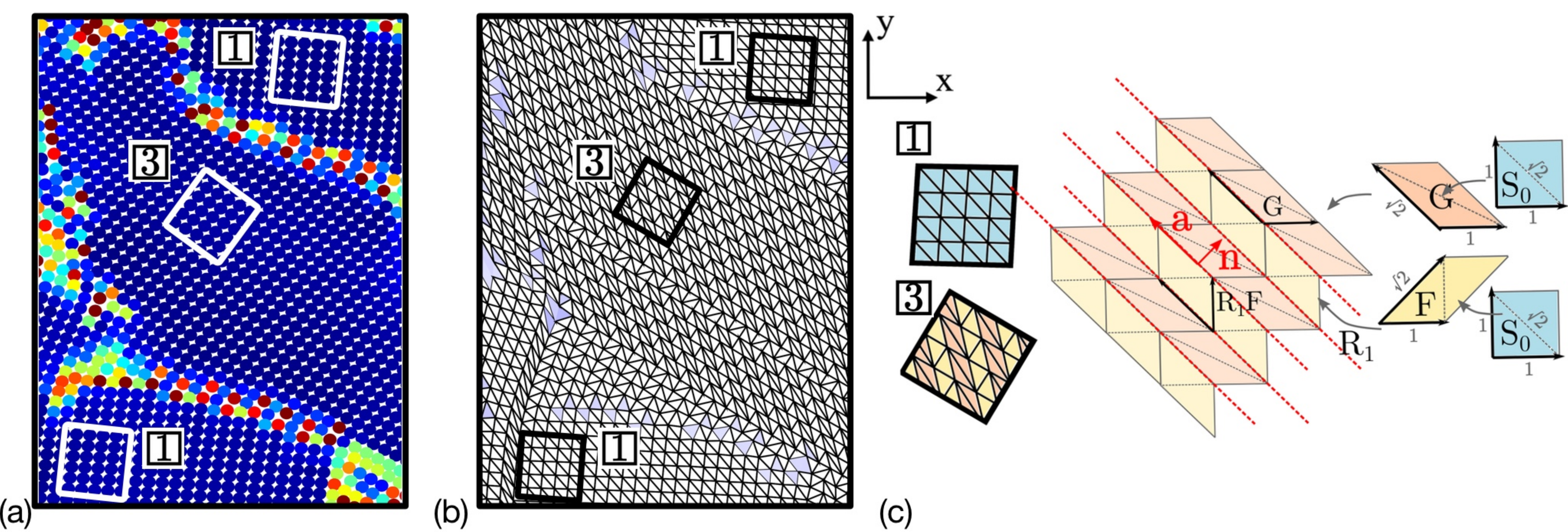}
\caption{\footnotesize {A magnified fragment  of the post-avalanche pattern from Fig.~\ref{fig:patt} showing the nodal coordinates in (a) and the actual affine distortions  of individual finite elements in (b).   The fragment is chosen to emphasize the coexistence of rotated patches of the type 1 and 3 and the micro-twin laminate structure of rotation in the patches of type 3. In (b) the   finite element triangles are colored   according to  the level of their  strain-energy   but only when it is above  a certain threshold: uncolored elements have  very low energy;
 (c)  illustration of the    solution   of the compatibility equation describing the laminate structure of the rotated patches of type 3. 
 } }
\label{fig:twinmode}
\end{figure*}

\begin{figure*}[hbt!]
\centering
\includegraphics[width=1.5\columnwidth]{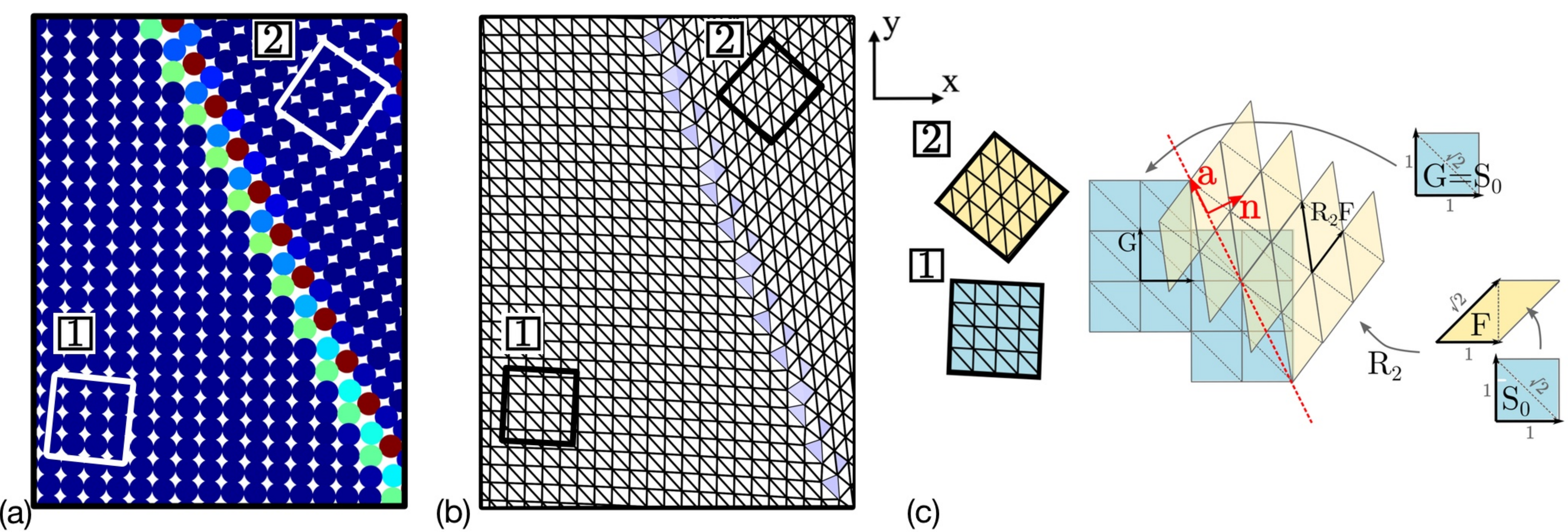}
\caption{\footnotesize {A magnified fragment  of the post-avalanche pattern from Fig.~\ref{fig:patt} showing the nodal coordinates in (a) and the actual affine distortions   of individual finite elements in (b).  The fragment is chosen to show the coexistence of patches of the type 1 and 2.   Finite element triangles in (b) are colored according to  the level of their  strain-energy  but only when it exceeds a certain threshold: uncolored traungles have low energy;
 (c)  illustration of the corresponding  solution   of the compatibility equation. 
 } }
\label{fig:twinmode2}
\end{figure*}

 In Fig.~\ref{fig:twinmode} and~\ref{fig:twinmode2},  we show in more detail two  representative fragments of the distorted network of  our elastic elements   illustrating two different types of interfaces between misaligned patches of the original lattice. The fragment,  presented in Fig.~\ref{fig:twinmode}(a,b,c)  details the rotation mechanism shown in Fig. \ref{fig:skinny_patt2} but now in the context of an actual  decomposition pattern from Fig. \ref{fig:patt}.  The observed microstructures can be predicted based on the strain compatibility requirements. 
 
\emph{ Geometric compatibility. }Indeed, consider the deformation field $\bf y(\bf x)$   whose deformation gradients $\bf F={\bf \nabla} \bf y _+$ and $\bf G={\bf \nabla} \bf y _-$  are discontinuous on the surface $\Sigma$. For the deformation itself to remain continuous on the surface $\Sigma$,   the matrices  $\bf F$, and $\bf G$ must be rank-one connected which constitutes the kinematic  (Hadamard) compatibility condition 
 \begin{equation}
 \mathbf{R}\mathbf{F}=\mathbf{G}+\mathbf{a}\otimes\mathbf{n}^*=\mathbf{G}\left(\mathbf{I}+\mathbf{a}^*\otimes\mathbf{n}^*\right) =\left(\mathbf{I}+\mathbf{a}\otimes\mathbf{n}\right){\bf G}, 
 \label{eq:twin_compatibility}
 \end{equation}  
where ${\bf R}\in SO(2)$ is a rotation. 

The Eulerian vector $\bf n$ (normal to the discontinuity plane) and covector $\bf a$, defining the amplitude of the shear, must satisfy  ${\bf a}\cdot{\bf n}=0$; their Lagrangian counterparts are  $\mathbf{a}^*=\mathbf{G}^{-1}\mathbf{a}$ and  $\mathbf{n}^*=\mathbf{G}^T{\mathbf{n}}$.   
If we assume further that $\det \mathbf{F}=\det \mathbf{G}=1$ and exclude reflections,  the deformation gradients satisfying (\ref{eq:twin_compatibility}) form a  mechanical twin.  If, in addition,  the rotation $\mathbf{R}$ belongs to the point group of the lattice, such a twinning structure produces the undistorted zero energy configuration.  The resulting micro-twinned laminates  are sometimes referred to as pseudotwins  \cite{Pitteri2002-rm}.

\emph{Twinning equation.}  The   equation \eqref{eq:twin_compatibility} was studied extensively, see for instance \cite{Ball1989-ba,Forclaz1999-hu}. It was shown   that (\ref{eq:twin_compatibility}) admits either no solutions  or two solutions. More specifically, 
the two solutions exist when the matrix  $\mathbf{G}^{-T}\mathbf{F}^T\mathbf{F}\mathbf{G}^{-1}\neq{\bf I}$ and its ordered eigenvalues $\mu_1<\mu_2$ are such that $\mu_1\mu_2=1$. In that case, the two solutions are given explicitly by the formulas:
\begin{align}
\mathbf{a}&=\rho\left(\sqrt{\frac{\mu_2(1-\mu_1)}{\mu_2-\mu_1}}\mathbf{v}_1+\kappa\sqrt{\frac{\mu_1(\mu_2-1)}{\mu_2-\mu_1}}\mathbf{v}_2 \right)\;,\\
\mathbf{n}&=\frac{1}{\rho}\left(\frac{\sqrt{\mu_2}-\sqrt{\mu_1}}{\sqrt{\mu_2-\mu_1}}\right)\left(-\sqrt{1-\mu_1}\mathbf{v}_1 +\kappa\sqrt{\mu_2-1}\mathbf{v}_2 \right)\;,
\end{align}
where $\mathbf{\hat v}_1$ and $\mathbf{\hat v}_2$ are the normalized eigenvectors of $\mathbf{G}^{-T}\mathbf{F}^T\mathbf{F}\mathbf{G}^{-1}$, $\rho>0$ is a constant ensuring that $|{\bf n}|=1$ and $\kappa=\pm 1$. Once  $\bf a$ and $\bf n$ are known, the rotation $\bf R$ can be obtained directly from \eqref{eq:twin_compatibility}. 

\begin{figure}[h!]
\includegraphics[width=.7\columnwidth]{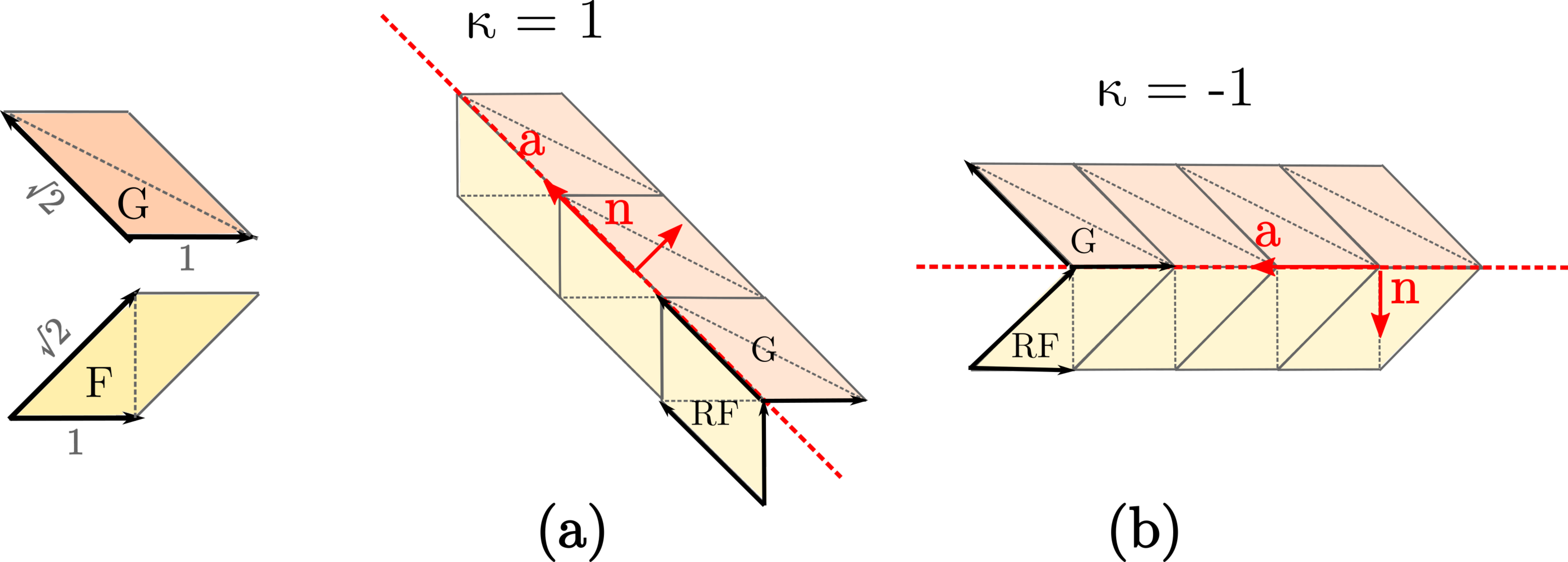}
\caption{The two solutions of the twinning equation \eqref{eq:twin_compatibility} for  the deformation gradients ${\bf F}={\bf S}^0_1$ and ${\bf G}={\bf S}^0_{-1}$.
}\label{fig:twinning}
\end{figure}

Here, we are interested in the special case  of  compatibility between two $GL(2,\mathbb{Z})$ related deformation gradients, in other words, for the case  when   $\bf G$ and $\bf F$ are two equivalent  minima of  the strain-energy $\phi$. 

Consider first pseudotwins shown in Fig. \ref{fig:skinny_patt2}  and Fig. \ref{fig:twinmode}(a,b),  where    $\mathbf{F}$ is in ${\bf S}^0_{1}$ and  $\mathbf{G}$ is in $\mathbf{S}^0_{-1}$.
For this  case the condition $\mathbf{G}^{-T}\mathbf{F}^T\mathbf{F}\mathbf{G}^{-1}\neq{\bf I}$ is satisfied 
and  the solution corresponding to  $\kappa=1$  is the one observed in our  post-instability patchy pattern. It is characterized by the parameters
$
\mathbf{a}^T=\lbrace -\sqrt{2},\sqrt{2} \rbrace,  \mathbf{n}^T=\lbrace \cos \zeta, \sin\zeta \rbrace, \zeta=\pi/4
$
with $\mathbf{R}$ a counter-clockwise rotation of $\psi=2\zeta=\pi/2$, see our Fig. \ref{fig:twinmode}(c) and Fig. \ref{fig:twinning}(a). 
For $\kappa=-1$, the solution is  
$
\mathbf{a}^T=\lbrace -2, 0 \rbrace, \mathbf{n}^T=\lbrace \cos \zeta, \sin\zeta \rbrace,  \zeta=-\pi/2
$
with $\mathbf{R}(\psi=0)$. The corresponding micro-twins, see Fig. \ref{fig:twinning}(b),  can be seen, for instance,  in Fig. \ref{fig:twinmode}(b) in the bottom left 'triangle' between the rotated patches of the type 1 and 3. Note that such grain boundaries are neither microscopically nor macroscopically compatible as the corresponding habit planes do  not exist in this system. The study of their dislocational structure will be presented separately.


%
%
%
%
%

%
  
\begin{figure}[h!]
\includegraphics[width=.7\columnwidth]{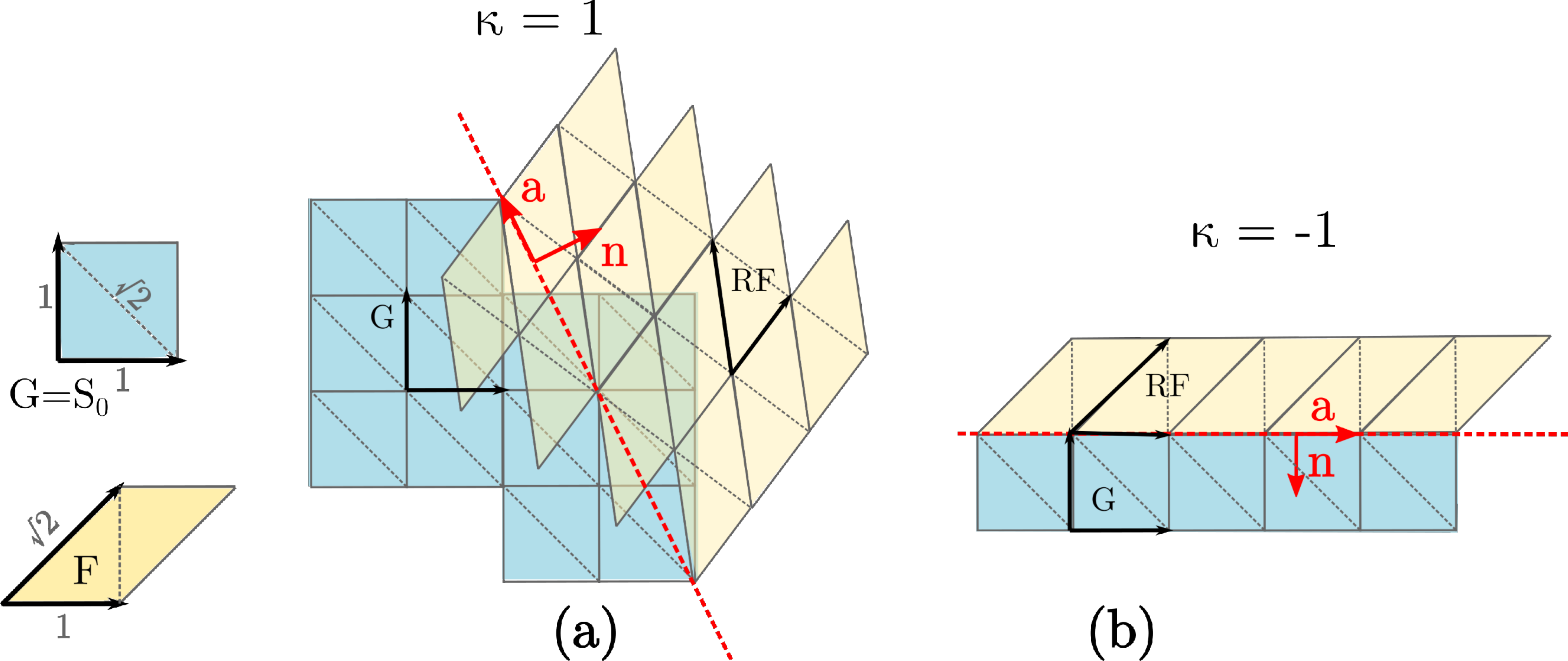}
\caption{The two solutions of the twinning equation \eqref{eq:twin_compatibility} for  the deformation gradients ${\bf F}={\bf S}^0_1$ and ${\bf G}={\bf S}^0=\textbf{I} $.
}\label{fig:twinning2}
\end{figure} 
A more conventional case of misoriented  coexisting patches of the original lattice (patches of the type 1-2) is shown in Fig. \ref{fig:twinmode2}(a,b). Here the    deformation gradients ${\bf G}$ and $ {\bf F}$   correspond to the bottoms of the energy wells ${\bf S}^0$ and ${\bf S}^0_1$, respectively. 
The two solutions of the  twinning equation  are: $\mathbf{a}^T=\lbrace -\sin\zeta, \cos\zeta  \rbrace ,\mathbf{n}^T=\lbrace \cos\zeta, \sin\zeta\rbrace, \mathbf{R}(\psi=2\zeta), \zeta=\arctan(1/2)$, for  $\kappa=1$,  and  $\mathbf{a}^T=\lbrace 1,  0  \rbrace, \mathbf{n}^T=\lbrace \cos\zeta,\sin\zeta \rbrace, \mathbf{R}(\psi=0),  \zeta=-\pi/2$,  for $\kappa=-1$. The solution  corresponding to $\kappa=1$ was observed  in our post-avalanche patchy pattern, see our  Fig. \ref{fig:twinmode2}(b), but not as a   microscopically compatible  micro-twin laminate  but,  instead, as a macroscopically compatible but microscopically   semi-coherent low-energy interface  known as $\Sigma 5$ grain boundary  \cite{Gao2020-qy}, see Fig. \ref{fig:twinmode2}(c) and Fig. \ref{fig:twinning2}(a). The solution corresponding to $\kappa=-1$, Fig. \ref{fig:twinning2}(b),  was not observed as a macroscopic grain boundary even though  the associated configuration  was effectively developing each time a dislocation was crossing  the crystal. 
\begin{figure*}[hbt!]
\centering
\includegraphics[width= 1.2\columnwidth]{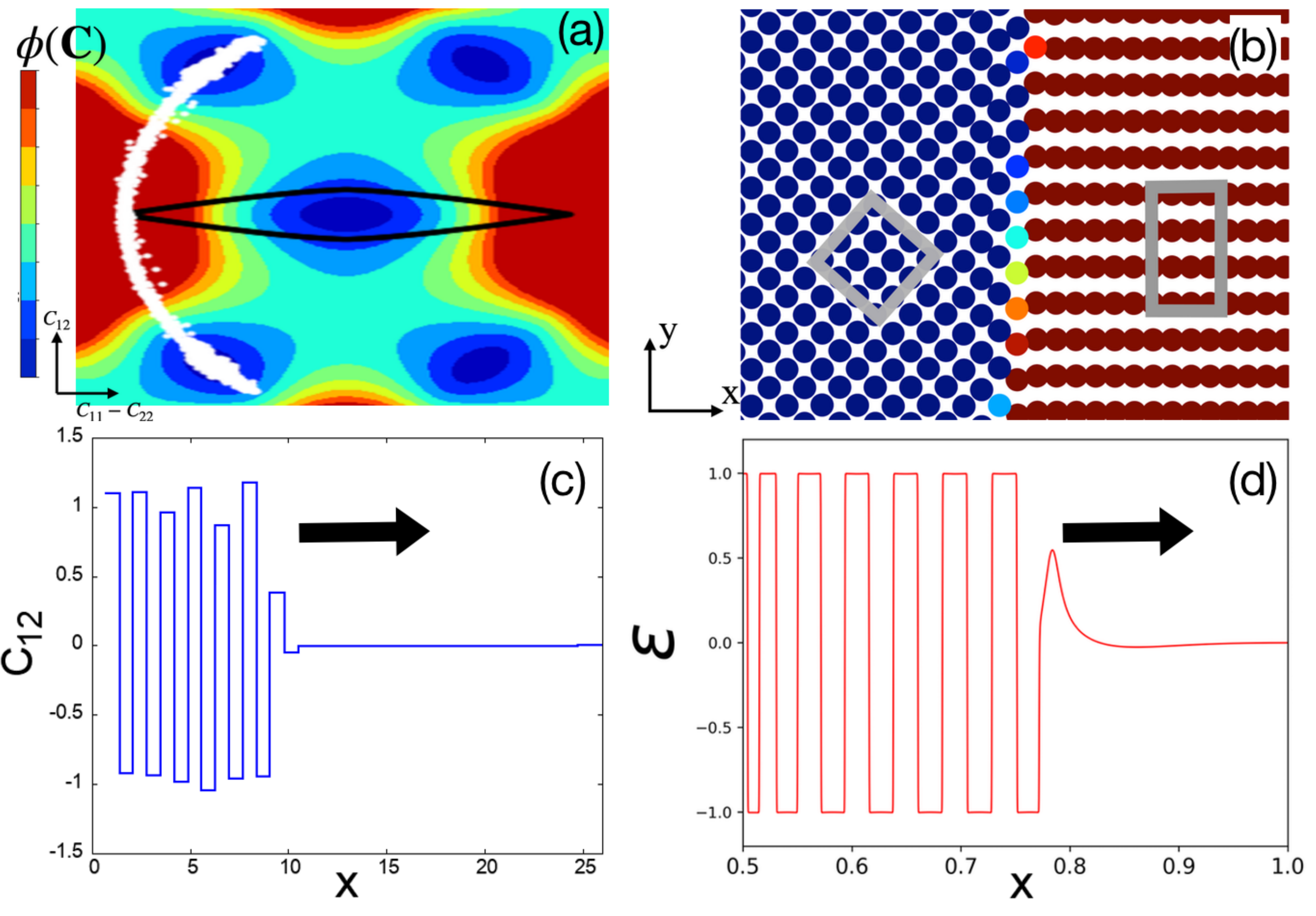}
 \caption{\footnotesize {The unfolding of the dislocation avalanche in the  single-slip-biased version of the model.    (a) Energy landscape showing the post-avalanche spreading of configurational points; (b) the structure of the moving front separating the initially stretched rectangular lattice and the growing micro-twinned  (rotated) square lattice (colors indicate the level of energy); (c) the distribution of the shear strain $C_{12}$ across the transformation front shown in (b);   (d)  propagating front of lamination in the   one dimensional toy model with parameters   $\beta=3\times10^{-6}$, $\eta=0.0017$, $\gamma=10^{-6}$}, $A=-0.001$.}
\label{fig:11}
\end{figure*}

 
Note that despite the full stress relaxation inside the two types of rotated patches shown in Figs.~\ref{fig:twinmode} and \ref{fig:twinmode2}, the associated rotations are neither elastic nor rigid. Instead, they are achieved through distributed crystallographically exact shear and are therefore slip-induced and fully dissipative, see also \cite{Ovidko2011-vw} for related experimental observations. In the next section, we elucidate the mechanism of such inelastic rotations by simulating analogous phenomena in the specially designed prototypical  models.

%

\section{Prototypical  models} 
 \label{sec:pm}
To accentuate the   mechanism of 'rotation by micro-twinning', shown Fig.~\ref{fig:twinmode}(a,b),  we consider below a prototypical energy density,  specially designed to  ensure   that  only one slip mechanism, involving the energy wells $ {\bf S}^0_1$ and ${\bf S}^0_{-1}$, is activated. 

\emph{2D prototypical model.} Such bias  can be achieved if, in particular,  the volumetric distortions are  strongly penalized.  The  prototypical  energy density  is    taken in the additive form 
\begin{equation}
\phi(\bC)=\phi_v(\det\bC)+\phi_s(\bf \tilde{C}),
\end{equation}
where $\bf \tilde{C} =\bC/(\det\bC)^{1/2}$. The   volumetric part $\phi_v(s) $ is  chosen in the form  $ \mu  (s-\log(s))$  with stiff bulk modulus $\mu=25$  which precludes (physically likely) softening in tension  while  still banning  configurations with infinite  compression.   

The volume preserving shear energy $\phi_s$  needs to be specified only inside a single  domain of periodicity and then extended by global symmetry.  The convenient choice is the lowest order  polynomial  which  ensures  the   continuity of the elastic moduli \cite{Parry1998-sv}. A minimal  expression of this type  was     proposed in \cite{Conti2004-sv} and we adopted it in our simulations. It has the form 
\begin{equation}
\phi_s({\bf \tilde{C}}) =  \beta \psi_1 ( {\bf \tilde{C}}) 
 + \psi_2 ({\bf \tilde{C}}),    
\end{equation}
where 
$
\psi_1={I_1}^4\,I_2 - 41\,{I_2}^3/99 +
7\,I_1\,I_2\,I_3/66 + {I_3}^2/1056, $ and $
\psi_2 = 4\,{I_2}^3/11  + {I_1}^3\,I_3 -  8\,I_1\,I_2\,I_3/11  +  17\,{I_3}^2/528. 
$
Here we used the  (hexagonal) invariant functions of the (normalized) metric tensor:
$
I_1 =  \frac{1}{3} (\tilde{C}_{11} + \tilde{C}_{22} - \tilde{C}_{12})$, 
$I_2= \frac{1}{4} (\tilde{C}_{11} - \tilde{C}_{22})^2 + \frac{1}{12}(\tilde{C}_{11} + \tilde{C}_{22} -
4 \tilde{C}_{12})^2$, and 
$I_3 =  (\tilde{C}_{11} - \tilde{C}_{22})^2 (\tilde{C}_{11} + \tilde{C}_{22} - 4 \tilde{C}_{12}) - \frac{1}{9} (\tilde{C}_{11} + \tilde{C}_{22} -
4 \tilde{C}_{12})^3.$  The choice  $\beta =- 1/4$   enforces the   square symmetry on the reference state.


In  Fig. \ref{fig:11}(a) we show  that under our loading protocol such  choice of the energy density indeed biases the system towards activating only one slip system. The important here is not the particular structure of the function $\phi_s$, which is largely controlled by the $\mathrm{GL}(2,\Z)$ symmetry, but rather the structure of the function $\phi_v$  which  is responsible for the highly elongated shape of the apparent 'yield surface',  see Fig. \ref{fig:11}(a). Indeed, with this configuration of the  'yield surface', the loading in pure shear $\mathbf{F}_\rectangle (\alpha)$, which means moving left from the origin along the horizontal path on the Poincare disk, leads to the breakdown in the immediate vicinity of the energy wells $ {\bf S}^0_1$ and ${\bf S}^0_{-1}$.  In Fig. \ref{fig:11}(b) we show that during the avalanche in such   system,  a transition front forms  separating the  (rectangular) stressed  configuration  $ {\bf S}$ and the unstressed stable laminate involving the states ${\bf R}(\pi/2){\bf S}^0_1$ and ${\bf S}^0_{-1}$. This laminate takes the form of an apparently rigid $\pi/4$ rotation which develops behind the propagating front.

The corresponding transient 'computational dynamics' inside the avalanche is illustrated in a movie presented in the  Supplementary Material.  More specifically, our \textbf{Movie S3}  shows the fast time evolution of the deformation field ${\bf y} ({\bf x})$  when we use in the numerical experiments the polynomial strain-energy density described above.  Compared  to what we saw in \textbf{Movie S1}  and \textbf{Movie S2},  here we observe a much more organized evolution that remains effectively 'laminar' throughout the whole avalanche. Once again the two fronts, separating affine and non-affine dynamics, start to propagate from the vertical boundaries of the body, however, in this artificially designed model they remain stable and the fast time 'computational flow' never develops any 'vortices'. Instead, we observe the fast side (transversal) motion of dislocations which is highly organized as it leaves behind a new homogeneous state.  In other words, a non-affine transient evolution results in an affine final configuration (modulo a single residual defect resulting from a computational noise). Since the lattice scale micro-twinning which is disguised here as rigid rotation is accomplished   through the side motion of dislocations,  the resulting deformation is inelastic and the corresponding rotation should be qualified as dissipative. 

\emph{1D prototypical model.} An even more schematic,  1D prototypical description of  the  propagating transition front,  imitating the one shown in Fig. \ref{fig:11}(b), can be obtained if we neglect the transversal motion of dislocations  and focus instead on the development of a laminate stabilized by competing interactions. To this end we need to introduce   a   potential  
\begin{equation}
\label{eqdyn1}
f(\epsilon)=(A/2)\epsilon^2-(1/4)\epsilon^4+(1/6)\epsilon^6 , 
\end{equation}
where $\epsilon=u_x$ and  $u(x)$ is a scalar displacement field. The energy density \eqref{eqdyn1} may have up to three  wells with the higher-symmetry state playing the role of the deformed configuration  $ {\bf S} $,  and the two lower-symmetry states representing the symmetric variants  $ {\bf S}^0_1$ and ${\bf S}^0_{-1}$.

 Consider next on the finite domain $0 \leqslant x \leqslant 1$  a continuum 1D dynamical system with the energy 
\begin{equation}
\label{eqdyn2}
E=\int_0^1[f(u_x) + \frac{\beta}{2}u_{xx}^2 + \frac{\gamma}{2}u^2 ]dx.
\end{equation}
Here the higher order second term $u_{xx}$ represents  the strain gradient  regularization 
and brings an internal length scale. The lower order term $u^2$  represents the energy of the constraining elastic environment and also  brings a (competing)  length scale into the problem. Introducing  the Rayleigh type dissipation~\cite{Lookman2003-xv}
\begin{equation}
\label{eqdyn3}
R=(\eta/2) \int _0^1\dot u_x^2dx,
\end{equation}
where $\eta$ is the  effective viscosity coefficient, we obtain the  dynamic  equation 
\begin{equation}
\label{eqdyn}
\eta u_{xxt}= -f''(u_x) u_{xx}+\beta u_{xxxx}+\gamma u ,
\end{equation}
whose role is to imitate the overdamped   'computational dynamics' operative inside the duration of the avalanche.

To solve the equation \eqref{eqdyn} numerically, we first approximate spatial derivatives by finite differences using a fixed grid with size $\Delta h$ and   utilize semi-implicit forward Euler discretization in time with a time step $\Delta t$. The resulting discrete set of equations in Fourier space takes the form 
\begin{equation}
\hat u^{t+\Delta t}(q) =  \frac{\eta\hat M_3(q)\hat u^{t}(q) - \Delta t   \hat M_1(q) \hat f'(q)}{   \eta \hat M_3(q) -  \text{d}t (\beta \hat M_2(q)+ \gamma)},
 \end{equation}
where $\hat M_1(q) = \di  \sin(q)/(2\Delta h)$, $\hat M_2(q) = 16\sin(q)^4/(\Delta h)^4$ and,  $\hat M_3(q) = (2\cos(q)-2)/(\Delta h)^2$.
The discrete Fourier transform on a unit grid is defined as $\hat u(q)  = N^{-1}\sum_\mathrm{i} u_\mathrm{i}e^{- \di q x }$ with $x=\mathrm{i}$ and $q = 2\pi \mathrm{k}/N$, where $\mathrm{i}=0,1,...,N-1$ and,  $\mathrm{k}=0,1,...,N-1$.  The nonlinear function $f'=\partial f/\partial u_x$ is first evaluated in real space and then Fourier transformed to obtain $\hat f'(q)$. The Fourier image of the corresponding displacement field is computed as $\hat u(q) = \hat \epsilon(q)/\hat M_1(q)$ for $q\neq0$.  In our simulations, we used the parameter values: $\eta=0.0017$, $\gamma=10^{-6}$, $A=-10^{-3}$, $\beta=3\times10^{-6}$, $ \Delta h =0.04$, $N=8192$, and  $\Delta t=0.004$. The initial data were chosen in the form a  localized strain increment $\epsilon(x) = 0.5 e^{-(x-x_m)^2}$  centered around  $x_m=0.5$. Similar processes of growth of a stable micro-laminate at the expense of an unstable homogeneous state have been studied before in many other settings \cite{Van_Saarloos2003-oz,Ren2000-le,Doelman2009-mt}.

\begin{figure}[hbt!]
 \centering
\includegraphics[scale=0.22]{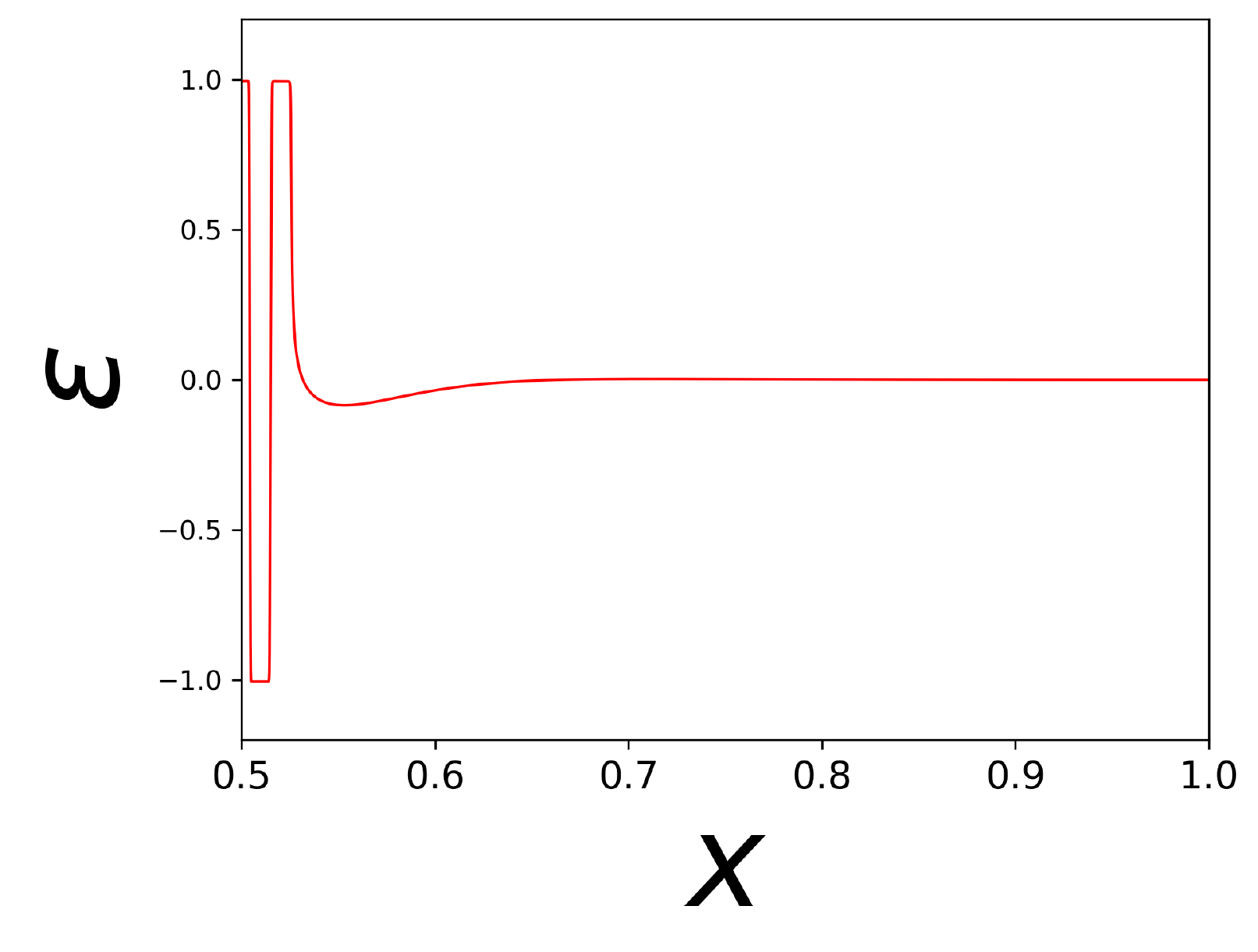}
\includegraphics[scale=0.22]{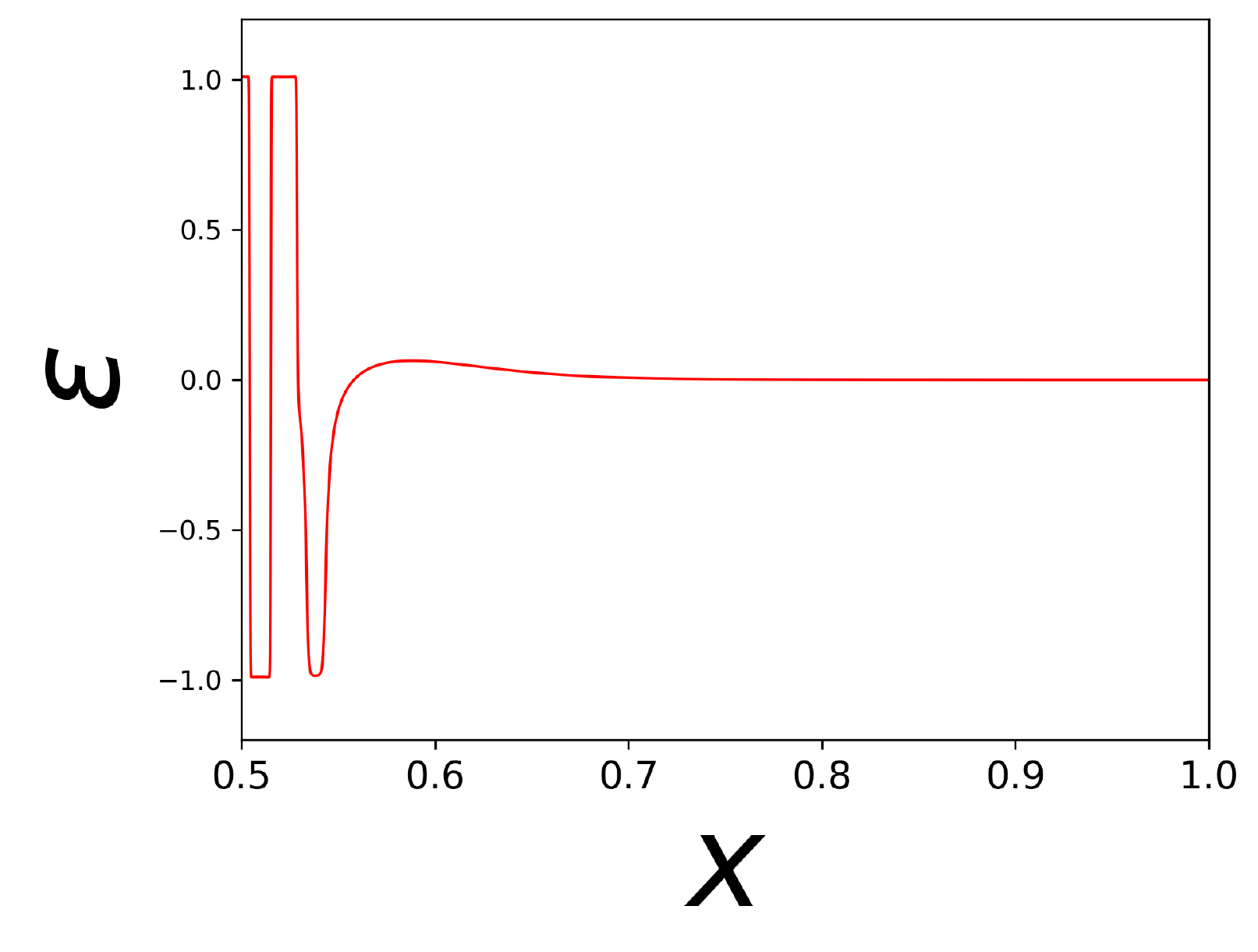}\\
\includegraphics[scale=0.22]{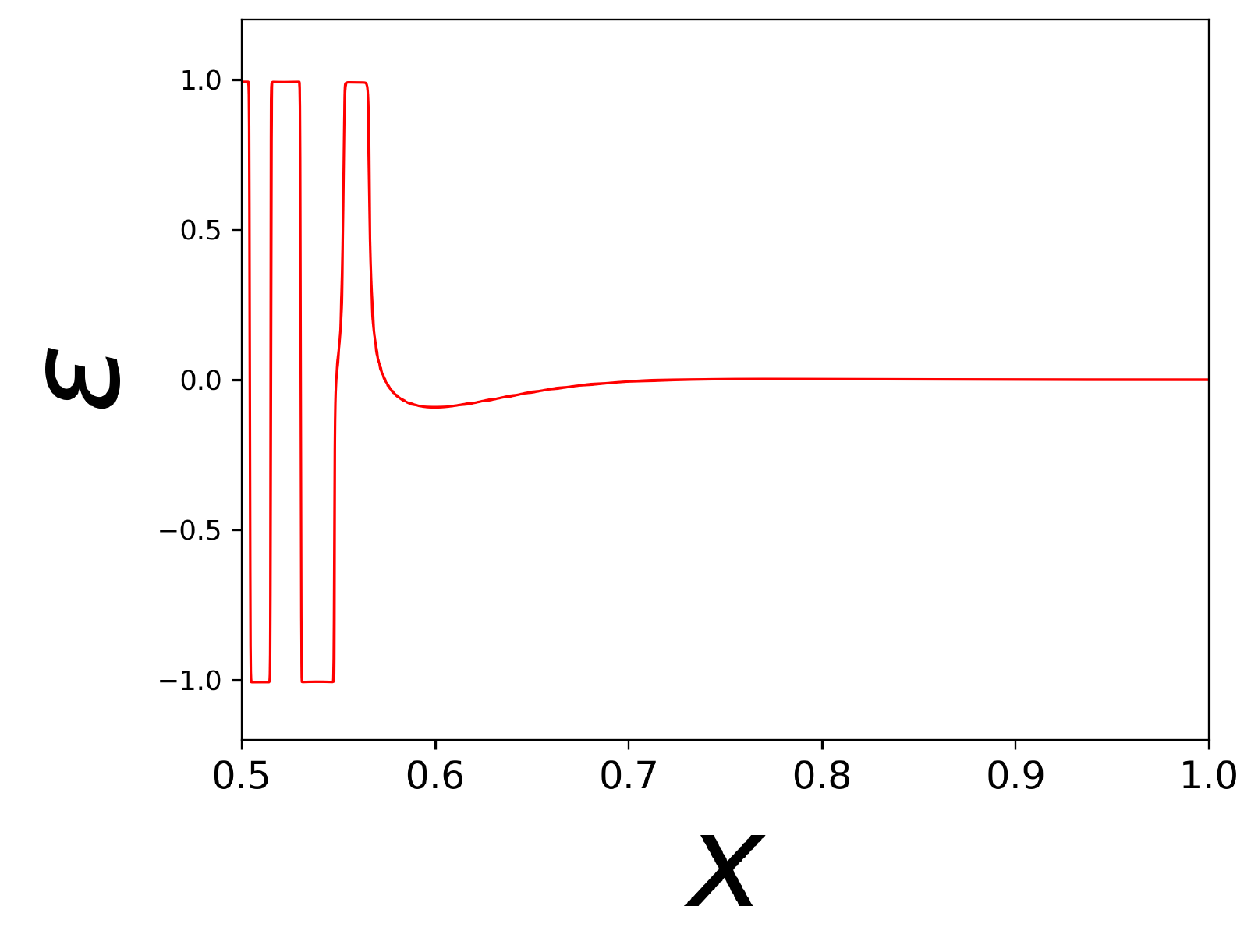}
\includegraphics[scale=0.22]{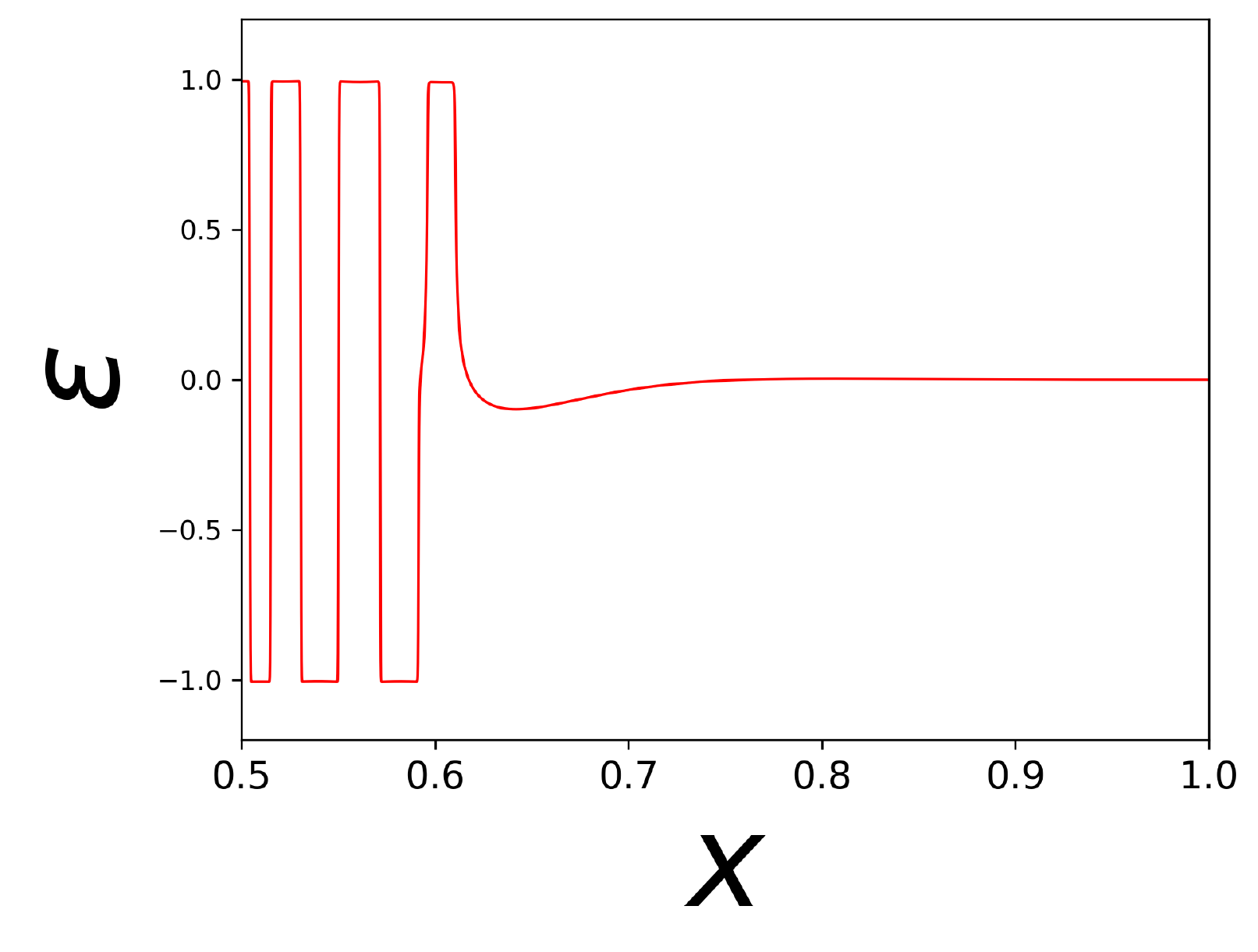}\\
\includegraphics[scale=0.22]{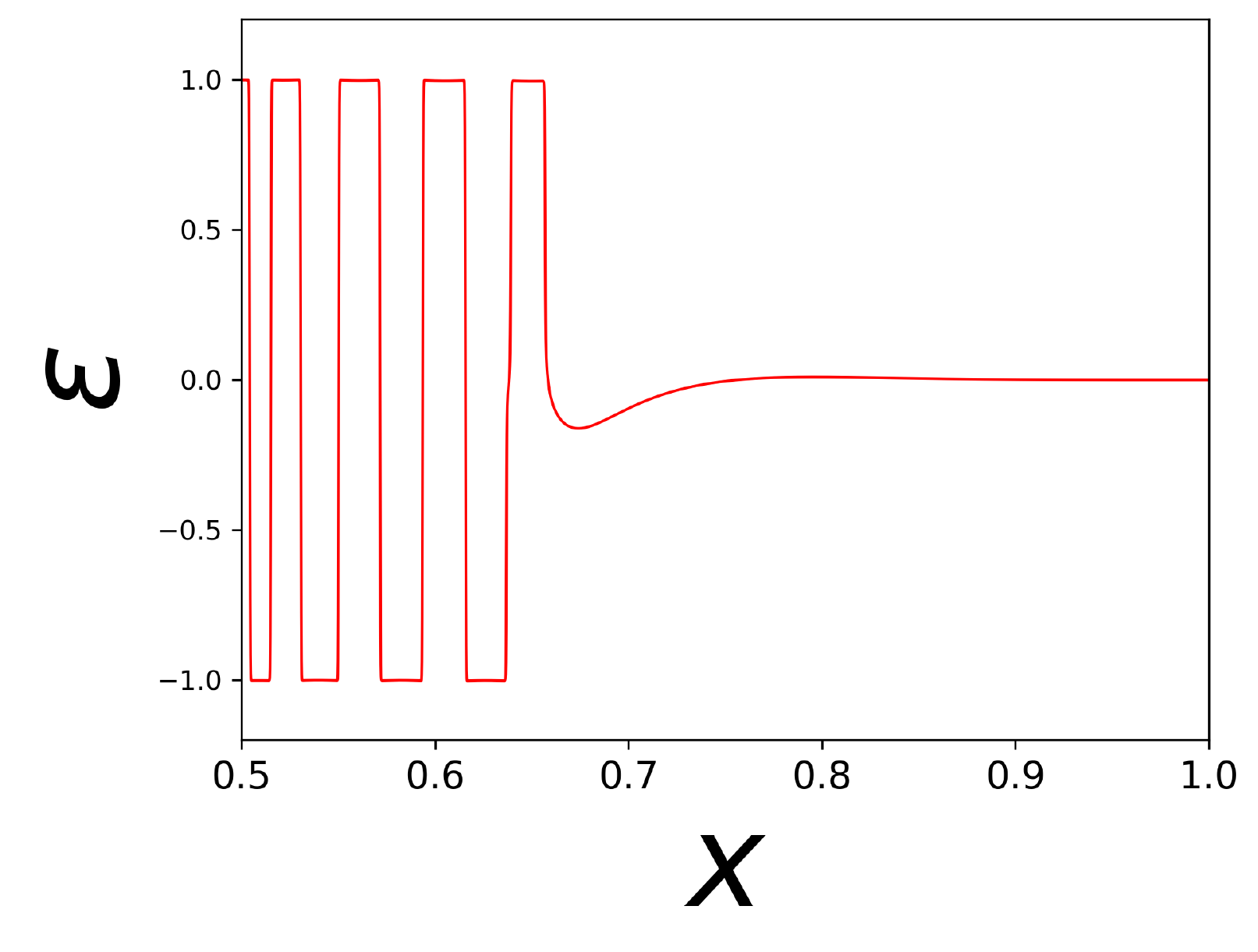}
\includegraphics[scale=0.22]{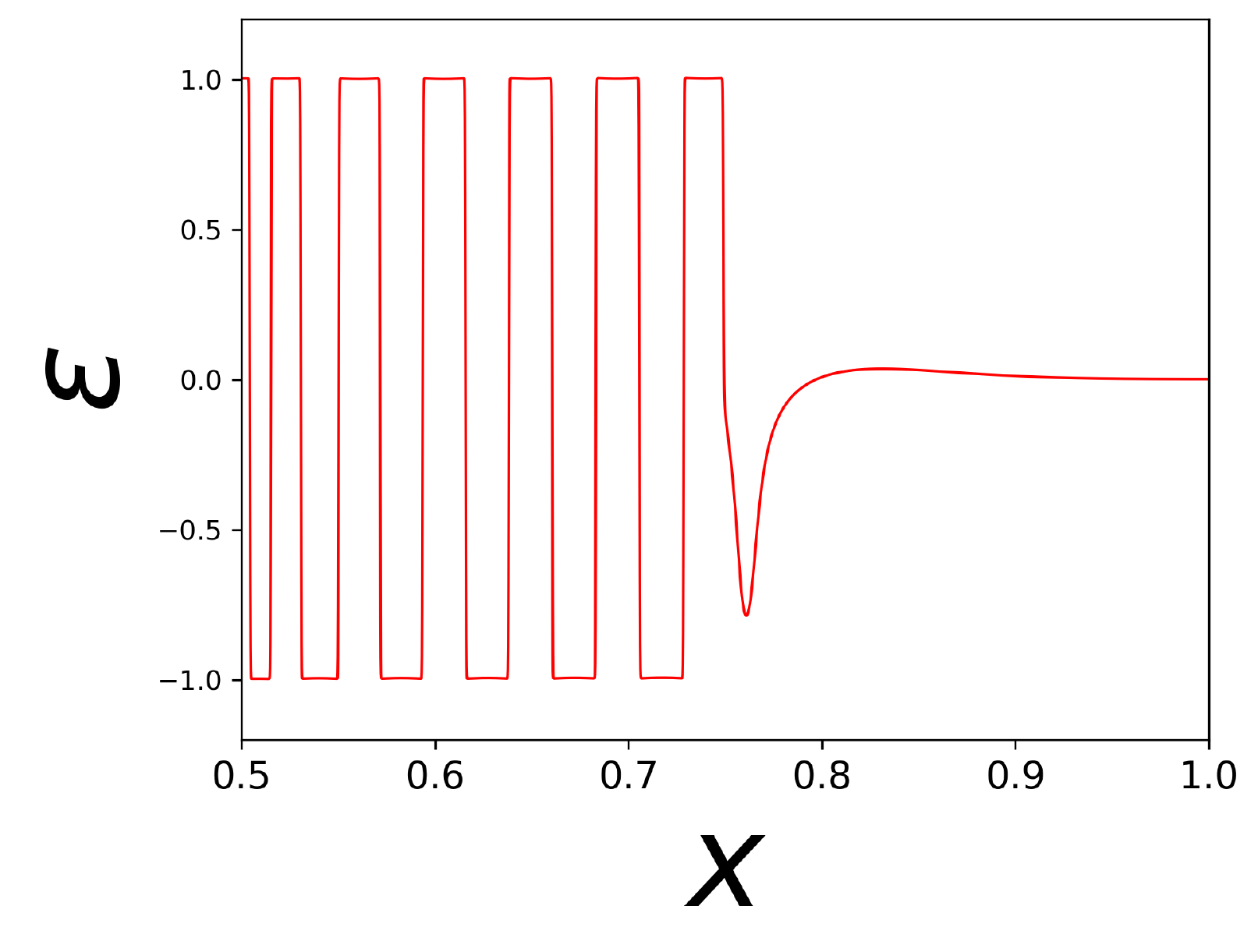}
\caption{Evolution of the  strain field  $\epsilon(x)$ at times $t=390, 480, 580, 760, 970, 1350\times \text{d}t$. The system ultimately evolves to the fully transformed state, which is a mixture of  $\epsilon(x)\approx\pm1$. }
\label{fig2_ap}
\end{figure}
  
  The  numerical solution of  Eq. \ref{eqdyn}   is compared   in Fig. \ref{fig:11}(c,d)  with the corresponding  inter-avalanche  dynamics in the   single-slip  prototypical version of our   2D model. The snapshots of the time  evolution   of the   strain field  $\epsilon(x,t)$  are shown in Fig. \ref{fig2_ap}. The actual dynamic development of the micro-twin pattern  can be also followed  in a  movie presented in the Supplementary Material. 
  
  More specifically, our  \textbf{Movie S4}    illustrates the dynamic growth of a laminate from a homogeneous state in front of it. An initial strain perturbation, which was used to initiate the dynamics,  was localized on one of the boundaries.  One can see that the pattern formation emerges as an invasion process, in which a piece-wise smooth  inhomogeneous state, which is stable,  takes over a marginally unstable homogeneous state.  We observe that the initial perturbation first grows in amplitude and then spreads in a form of a spatially periodic pattern.   The ensuing growth of a laminate appears macroscopically as a front,   traveling with a constant velocity,  and apparently winning over the growth of bulk modes.   The comparison of the propagating patterns in Fig. \ref{fig:11}(c,d) suggests that despite drastic simplifications, the 1D model captures the essential features of the corresponding dynamic process in 2D.

\section{Pseudo-turbulence}
\label{turb}
We have seen  that, in contrast  to the  intentionally over-simplified scenario described above, the actual 2D problem exhibits much higher complexity of the post-avalanche texture.  The access  in such a problem to a  broad variety of low energy configurations,  enabled  by inelastic rotations,   foments the development of imperfection-sensitive and therefore necessarily complex grain structure.  The post-instability 'fluidity' of the system, associated with the sudden dramatic   'opening'  of the energy landscape, makes the system-size plastic avalanche  inherently unstable. However, the system is still   under control  of   elastic interactions and the necessity to minimize elastic energy   leads to  self-organization. The latter is  manifested by the development of long-range correlations \cite{Gelin2016-sm,Geslin2021-qn}. 

\begin{figure}[h!]
\includegraphics[scale=0.15  ]{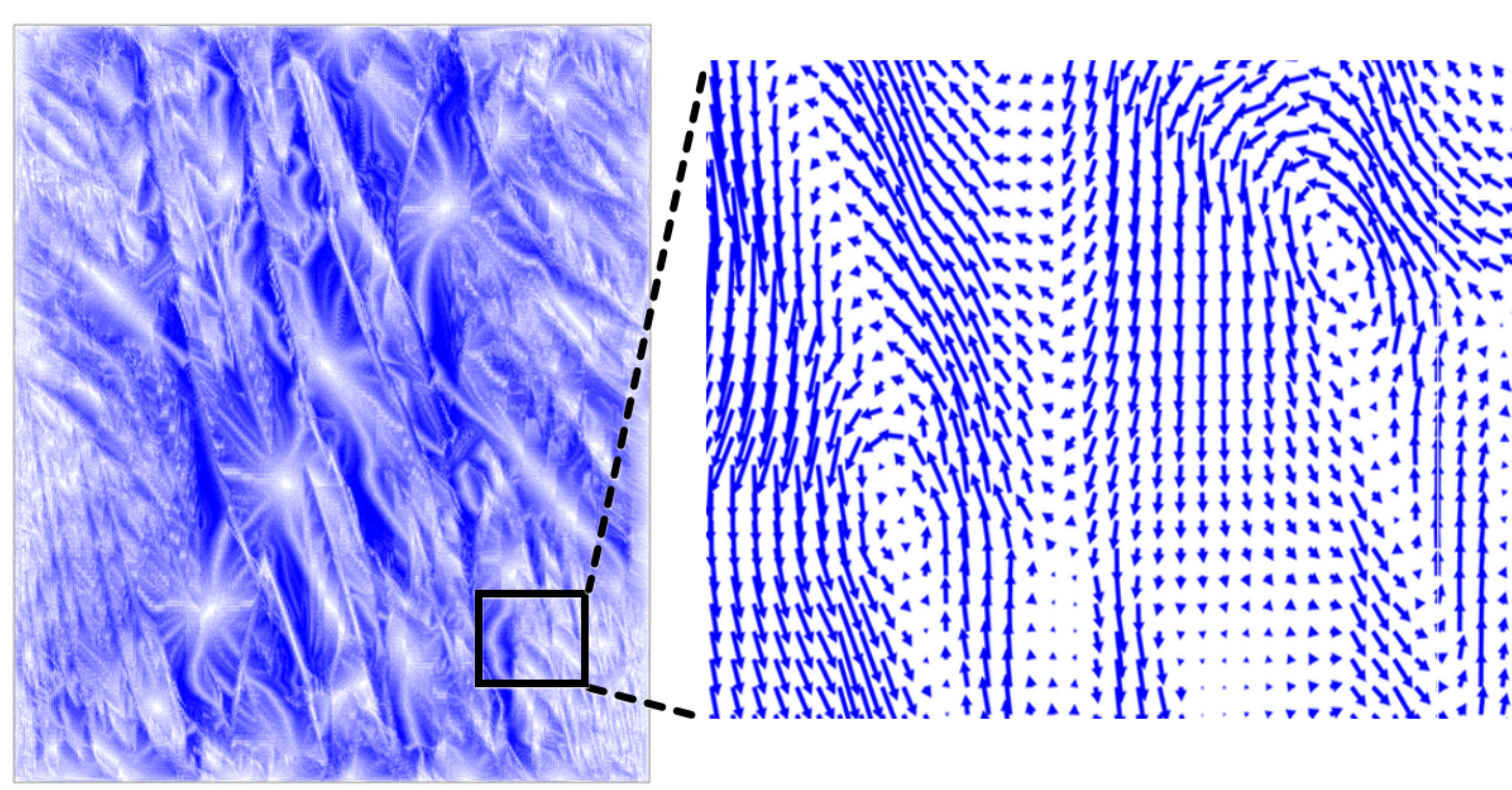}
\caption{\footnotesize{ Pseudo-turbulent   structure of the deformation field resulting from a dislocation avalanche.  The displacement vectors  connect pre- and post-avalanche  positions of  the individual nodes.
  }} \label{fig:pvslj211}
\end{figure}

To emphasize  the  intermittent  nature of the evolution of the inter-avalanche  deformation,   following a similar analysis of the granular system  \cite{Radjai2002-cj},   we generate a displacement field  connecting the dislocation-free state 'before' the avalanche and the dislocation-rich state 'immediately  after' the avalanche and study its spatial complexity. More specifically, we compute       the fluctuating part of the  total  nodal displacements   ${\bf u}({\bf x}) = {\bf y}({\bf x}) - {\bf F_\rectangle}(\alpha_c){\bf x}$. It is  shown   in Fig.~\ref{fig:pvslj211} with a zoom on two correlated turbulent-type  'eddies'. 

\begin{figure}[h!]
\centering
\includegraphics[scale=.073 ]{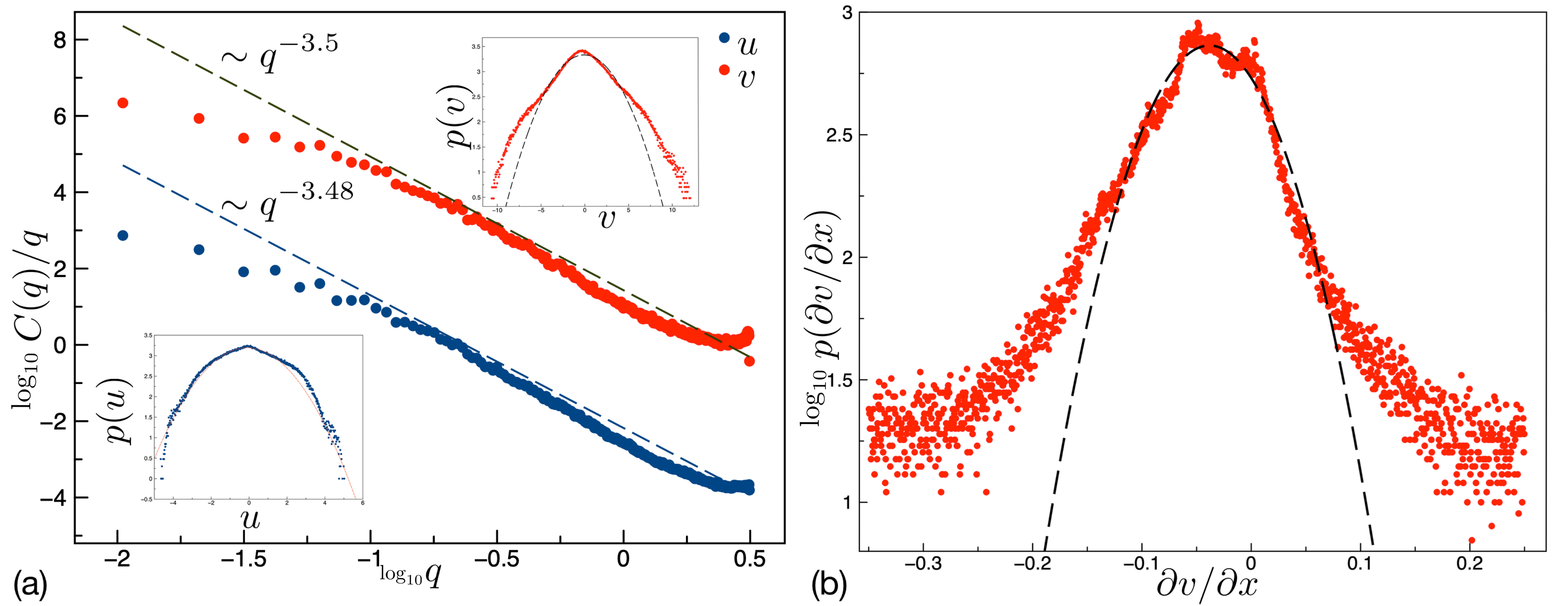}
\caption{\footnotesize{(a) Fourier transform of auto-correlation function (power spectrum density)
 of the  horizontal   $u$ and  vertical   $v$ components of the  displacement field; we interpret both components   as    surfaces  which due to averaging are transitionally invariant and isotropic;the dashed lines  show the    power-law fit; 
 insets show probability  distributions  $p(u)$ and $p(v)$.
 (b) Probability  distribution  $p(\partial v/\partial x)$ of the horizontal   derivative of the displacement field   exhibiting  non-Gaussian  wide tails.
 }}\label{fig:seven}
\end{figure}

Furthermore, in Fig. \ref{fig:seven}(a)   we show   the  power spectral density (PSD)  of the  field ${\bf u}(\bf x)$, illustrating the presence  of a hierarchy of scales and revealing the collective nature of the implied self-organization process.   We recall that the PSD  of a field $u(\bold r)$ is a Fourier transform of the correlation function, $C(\bold q)=\mathcal{F}[\int  u(\bold r_0)u(\bold r_0+\bold r)d\bold r_0]$ and  since it  depends on  wave-number and orientation, we performed the radial  averaging over orientations to obtain the function   $C(q)$,  with $q=|\bold q|$,  shown Fig. \ref{fig:seven}(a). 

Observe that outside the  small wave-number threshold  this function  exhibits a power law   asymptotics $C( q )/  q \sim    q ^{-2H-2}$.
 Our model produced the non-integer values for the Hurst exponent  $H\approx0.75$ and for the fractal dimension $D=3-H \approx2.25$, independently of the choice of the component. Such scaling indicates the self-affine (rough) nature of the surfaces representing the displacement field and is indicative of hierarchical organization \cite{Sreenivasan1991-hf,Hou1998-fu,Persson2005-sq,Turk2010-tt}.  
 Moreover, similar to what is known about fluid turbulence,  our  Fig. \ref{fig:seven}(b) also shows that the probability distribution function for at least one of the displacement derivatives is also characterized by robust algebraic tails.

%
 
Based on these observations, one can argue that the spatial correlations displayed by our MTM-based numerical tests display scaling characteristics that are indicative of turbulent-type dynamic rearrangements.    Temporal correlations, associated with plastic yield, are also known to be of self-affine nature and to exhibit power-law correlations and they are captured by even the simplest scalar    version of the  MTM \cite{Salman2011-ij,Zhang2020-ax}. Similar analogies with fluid turbulence have been  drawn in the analysis of several other complex disordered systems with domineering long-range interactions that are subjected to quasi-static driving,  from brittle cracking to active evolution of tissues \cite{Perig2010-ih,Angheluta2011-ba,Borja_da_Rocha2020-vf,Biswas2020-jw,Mandal2020-qo,Beygelzimer2021-bf}. In particular, behind the term 'granulence' are the  simulations and experiments recording for plastic avalanches in a collection of sheared granular particles non-affine  displacement fields   exhibiting the probability distribution functions with much wider tails than Gaussian  \cite{Radjai2002-cj,Oyama2019-ub,Goldenberg2007-ao,Combe2013-zf,Richefeu2012-ch,Miller2013-dw}.
For related observations in crystal plasticity simulations involving experiments and non-MTM modeling techniques, see \cite{Fressengeas2009-yd,Chevy2010-oa,Angheluta2012-ho,Beygelzimer2021-bf,Alava2014-dc}. 

%


 One may question the analogy between plastic avalanches and fluid turbulence based on the fact that in the former case we compare two static states, 'before' and 'after' the instability, while in the latter, the motion is continuous both in space and in time. Moreover in fluid turbulence    the emerging scale invariance is related to a balance between the energy that is continuously injected at large length scales and removed at small length scales while in plastic avalanches such energy cascades are not apparent. 
 
  To substantiate the  analogy, however,  it is helpful to   interpret a  plastic avalanche as a dynamic process that converts the potential energy accumulated in the metastable state 'before',  into either acoustic radiation or heat,  as the system transforms into a stable state 'after'.  Such a transformation process starts indeed  from a   static configuration but then it reaches a dynamic stage. As the initially stored energy is exhausted,  the dynamics itself experiences  a  decay, so  eventually the system reaches a new static configuration. In this picture, the dynamic stage is clearly present, as well the energy cascade from large scales (stored elastic energy) to small scales (acoustic radiation and heat).  Although the associated dynamic process is not inertial, as in fluid turbulence, it can still generate  self-induced complexity due to severe nonlinearity and the dominance of long-range interactions. 
 
It is interesting that, while the physical nature of the  'pseudo-turbulence' in our overdamped solid system is clearly different from the classical inertial turbulence in fluids, the complexity-generating nonlinearity may be similarly \emph{geometric}, meaning \emph{universal}.  Thus, the first source of nonlinearity in MTM is simply  quadratic, as it is  associated with the necessity to use   finite strains in the  description of deformations. This type of geometric nonlinearity is unavoidable if one wishes to capture correctly large crystal rotations \cite{Grabovsky2007-ll,Finel2010-zw, Salman2019-cg}. While the second source of nonlinearity is of constitutive nature (multi-well Landau potential),  it can be also considered as geometric.     Indeed, it is related to the geometrical location of the energy wells which is fully controlled by the structure of lattice invariant shears.  The energy wells themselves can be simply parabolic as we have previously shown using the scalar version of the MTM  \cite{Salman2011-ij,Salman2012-oa}.  Moreover, similar to the case of high Reynolds fluid turbulence,  the evolution generated by MTM  is fully conservative outside intermittent avalanches.  Furthermore, due to the quasi-static nature of the loading, the avalanches themselves remain dissipative even as the microscopic viscosity (encapsulated in numerical energy minimization) tends to zero \cite{Puglisi2005-lg}. Finally, as in fluid turbulence, the governing equations of the MTM   become scale-free in the limit when the cut-off length   tends to zero which ultimately explains the emergence of scale-free behavior.  

\section{Conclusions}
\label{sec:conc}
Plastic deformations in crystals originate from the availability of soft deformation modes which are usually represented as combinations of lattice-specific slips and rigid rotations.  Above certain level of stress, elastic deformations are comperatively disfavored due to their high energetic cost. As a result,  plastically deforming crystals typically form cell-based textures with elastic energy localized only on the boundaries separating randomly oriented patches or grains that are rotated relative  to each other while being effectively unstressed. 

In this paper, we used a  novel  mesoscopic tensorial model (MTM) to show that even   large lattice rotations constituting such textures,  can originate from a highly coordinated inelastic slip at the microscale. More specifically,  we presented numerical evidence that behind such rotations may be crystallographically exact micro-slip laminates of a pseudo-twin type. 

The formation of such laminates can be viewed as an effective internal 'wrinkling'  of the crystal lattice and since a lattice independent   internal scale  is missing, the laminate microstructure forms at the scale of individual elastic elements.  While the micro-laminates disguise themselves  as  elastically neutral  rotations,  our numerical experiments reveal that the process of their formation is inherently dissipative. By studying the dynamic mechanisms leading to such rotations,  we showed that they  emerge from dislocation-mediated processes at the microscale.  

The new insights were obtained due to the ability of the MTM approach to capture not only dislocation motion but also the dynamics of individual slips. An important finding revealed by our numerical experiments is that dislocation avalanches, resulting in the formation of rotated patches,   involve pseudo-turbulent rearrangements and lead to deformation fields exhibiting power-law distributed spatial correlations.  




Being meso-scopic,  the MTM approach  is   necessarily a result of coarse-graining of an atomistic model,  but at a much finer scale than the classical continuum theory.  Therefore in MTM  the fully non-local atom-by-atom interactions are inevitably lost and, for instance,    dislocation core structures appear  as  blurred. However, as we have shown in our study, the   MTM is  capable to represent  adequately not only the long-range but also  the short-range interactions  of dislocations which are  essential  for  the description of dislocation nucleation and  annihilation  as well as the formation of their  stable  interlocked configurations.

The  MTM  approach should  be developed further to address some puzzling problems in materials science and crystal physics. One of them is to rationalize the  size dependence of the mechanical response of sub-micron structural elements, in particular, to explain their enhanced strength,   the erratic nature of their stress-strain curves, and their propensity to discontinuous yield and brittle failure.  A related challenge is to move from  the study of single avalanches to numerical experiments addressing   temporal intermittency in steady plastic flows. The goal here is to obtain fundamental insights into the origin of scale-free correlations in plastic fluctuations and to learn how to identify and interpret their statistical structure. 

In order to simulate   realistic crystalline structures, such as FCC, BCC, and HCP,  it will be necessary to extend the MTM approach from 2D to   3D   which should start with the construction of an appropriate  $GL(3,\mathbb{Z})$-invariant energy. The 3D extension of MTM can be used to advance from basic modeling to actual engineering applications, in particular,   for calibrating the augmented engineering continuum models accounting for microscopic effects \cite{Weiss2015-eh}. In particular, such development will equip engineering models with the tools to adequately reproduce intermittent mechanical  noise  and will open them to the development of fluctuation-based nondestructive diagnostic techniques.

\section{Acknowledgments} 
 O. U. S. and R.B. were supported  by  the grants ANR-18-CE42-0017-03, ANR-19-CE08-0010-01, ANR-20-CE91-0010,  L. T. by the grant ANR-10-IDEX-0001-02 PSL.


\end{document}